\documentclass[11pt]{article}
\usepackage{amssymb}
\usepackage{wasysym}
\newcommand{\be}{\begin{eqnarray}}\newcommand{\beq}{\begin{equation}}
\newcommand{\ee}{\end{eqnarray}}\newcommand{\eeq}{\end{equation}}
\newcommand{\eps}{\varepsilon}
\newcommand{\De}{\Delta}
\usepackage{graphicx} 
\title
{
Probabilistic approach to the length-scale dependence of the effect of water hydrogen bonding on 
hydrophobic hydration 
}
\author{Y. S. Djikaev\thanks{
Corresponding author. 
E-mail: idjikaev@buffalo.edu
}
\hspace{0.1cm} and \hspace{0.1cm}E. Ruckenstein$^{}$
\thanks{E-mail: feaeliru@buffalo.edu }
\hspace{0.2cm}  \\ 
\\ Department of Chemical and Biological  Engineering, SUNY at Buffalo, \\ 
Buffalo, New York  14260 }
\date{ \hfill }
\renewcommand{\baselinestretch}{2}
\begin{document}
\renewcommand{\baselinestretch}{1}
\maketitle
\renewcommand{\baselinestretch}{1}
\vspace{-0.6cm}   
{\bf Abstract.} 
{\small 

We present a probabilistic approach to water-water hydrogen bonding that allows one  to obtain an analytic
expression for the  number of bonds per water molecule as a function of both its distance to a hydrophobic
particle and
hydrophobe radius. This approach can be used in the density functional theory (DFT) and  computer simulations 
to examine particle size effects on the hydration of particles and on their  solvent-mediated  interaction. For
example, it allows one to  explicitly identify a water hydrogen bond contribution to the external potential
whereto a water molecule is subjected  near a hydrophobe. 
The DFT implementation of the model predicts  
the hydration free energy per unit area of a spherical hydrophobe to be sharply sensitive to the hydropobe 
radius for small radii and weakly sensitive thereto for large ones; this corroborates 
the vision of the hydration of small and large length-scale particles as occurring via different  mechanisms. 
On the other hand, the model predicts that the hydration of even {\em apolar } particles  of small enough
radii  may become thermodynamically favorable owing to the interplay of the energies of pairwise (dispersion)
water-water and water-hydrophobe interactions.
This  sheds light on previous counterintuitive observations (both
theoretical and simulational) that two inert gas molecules would prefer to  form a solvent-separated pair
rather than a contact one. 

}

\newpage 
\section{Introduction}
\renewcommand{\baselinestretch}{2}
A particle whereof the accommodation in water is accompanied by an increase in an associated free
energy is called ``hydrophobic" and is often referred to as a hydrophobe. The thermodynamically
unfavorable dissolution of a hydrophobe (whether microscopic or macroscopic) is  hydrophobic
hydration; the corresponding free increase results from  structural (and possibly energetic) changes
in water around the hydrophobe. The total volume of water affected by two  hydrophobes is smaller
when they are close together than when they far away from each other.  This gives rise to an
effective, solvent-mediated attraction between them which is also referred to as hydrophobic
attraction.

Hydrophobic effects (hydration and attraction) play a crucial role in various physical, chemical,
and  biological phenomena.$^{1-4}$ They also play an important role in the 
formation, stability, and unfolding of the 
native structure of a biologically active protein which constitute the core of  two most exciting
(and intrinsically related) topics of modern biophysics, namely, ``protein folding" and ``protein
denaturation",$^{5,6}$  although an assortment of interactions (including those of hydrophilic character$^{7}$) 
would most likely determine the driving force of these amazing phenomena.

Various mechanisms have been suggested to understand hydrophobic effects at a fundamental level and
develop a general theory of  hydrophobicity.$^{8-11}$ Virtually all theoretical models involve the
hydrogen bonding ability of water as a key element. 

The structure of liquid water, its dependence on the external conditions, and the role of structural
changes in hydrophobic phenomena have long been the subject of intense research. For ambient
conditions, it was first described as a locally ordered tetrahedral network of water
molecules.$^{12}$ Various anomalous properties of water (such as the density maximum
at 4$^{\circ}$ C at atmospheric pressure,  local maximum and minimum of the isobaric heat capacity at
constant pressure, etc...) are attributed to the ability of its molecules to form hydrogen bonds,
strong directional bonds with energy  much larger than the thermal energy $k_BT$. As an example of
water structure effects on hydrophobic phenomena, the  biological activity of proteins appears to
depend on the formation, maintenance, and breakup of a 2-D hydrogen-bonded network spanning most of the protein surface and
connecting all the surface hydrogen-bonded water clusters.$^{13}$ 

Although much experimental, computational, and theoretical research has been carried out, many
thermodynamic and molecular aspects of hydration remain to be clarified. In the positive
(unfavorable) free energy of dissolving a hydrophobe, the positive entropic contribution (due to the 
negative entropy change) dominates over the enthalpic contribution at room temperatures. 
The total contribution of hydrogen bonding to the hydration enthalpy depends both on the single bond energy
and the number of bonds that a water molecule can form in the hydrophobe vicinity and in the bulk.  The
propensity of water to form all possible hydrogen bonds, on one hand, and the constraint, imposed by a
hydrophobe on the configurational space available to vicinal water molecules, on the other hand, lead to a
large entropic cost. Some investigations have suggested that water is more structured near a hydrophobe, with
water-water hydrogen bonds both labile and stronger than in bulk, but several experimental and theoretical
studies have reported the opposite results. Despite remaining controversies,  the dependence of hydrophobic 
phenomena on the length scales of solute particles is considered to be proven.$^{14-16}$  

The hydration of small hydrophobic molecules (of sizes comparable to a water molecule) is believed to
be entropically ``driven" (and so is  their solvent-mediated interaction).$^{10,11}$  Such molecules
can  fit into the water hydrogen-bond network  without destroying any bonds. While this  results in a
negligible enthalpy of  hydration, the solute  constrains some degrees of freedom of neighboring
water molecules which    gives rise to negative hydration entropy and hence to positive hydration
free energy. However, such a simple mechanism has recently come under scrutiny$^{10,11,14,15}$ 
because  there are simulations$^{17,18}$ and theory$^{19}$ suggesting that, under some conditions,
the hydration of small hydrophobic molecules could be entropically favorable. 

The hydration of large hydrophobic particles is believed to occur via a different  mechanism.$^{10,11,19,20}$
When inserted into liquid water, a large hydrophobe breaks some hydrogen bonds in
its vicinity.  This would result in large positive hydration  enthalpy and hence in a 
free energy change proportional to the solute surface area (as opposed to being proportional to
the solute volume for small hydrophobes). Thus, the hydration of large hydrophobic particles is 
expected to be enthalpically driven (and so is their solvent-mediated interaction). 

As the thermodynamics of hydration is expected to change gradually from entropic for small solutes to
enthalpic for large solutes, so are the structural properties of liquid water in the vicinity of the 
solutes. It was argued$^{21}$ that if the solute-water attraction is sufficiently weak, there
may exist  a thin film of water vapor near large hydrophobic solutes but not small ones.  This generated much
controversy.$^{10,11,18,20,22}$ 

Hereafter we present a model for water-water hydrogen bonding that allows one to obtain an analytic expression
for the  number of bonds per water molecule as a function of both its distance to a hydrophobe and hydrophobe
radius. This function can be used in the density functional theory (DFT) and  computer simulations (either
Monte Carlo or Molecular Dynamics) to examine particle size effects on the hydration of particles  
and on their solvent-mediated  interactions over the entire small-to-large length-scale range. 

Note that we do not investigate drying or wetting transitions$^{23}$ as such;  once the accomodation of the
solute particle in liquid water occurred, the fluid density distribution  in the vicinity of the solute and in
the entire system is not subject to any "transformations".  The latter may be induced only by changes in
external thermodynamic variables  (either temperature or pressure or chemical potential).  We will consider a
"static" version of the hydration phenomenon, wherein the state of the system does not change after hydration
occurred.  

\section{The number of hydrogen bonds per water molecule near a spherical hydrophobic surface} 

Consider a spherical hydrophobic  particle of radius $R$ immersed in liquid water (Figure 1).  Even if one
assumes that the intrinsic hydrogen bonding ability of a water molecule is not affected by the  hydrophobe, in
its vicinity  a ``boundary" water molecule  forms a smaller number  of bonds than in bulk  because the surface
restricts the configurational space  available to other water molecules  necessary for a  boundary water
molecule to form hydrogen bonds. The  probabilistic model allows one to obtain an analytic expression for the
average number of bonds that a boundary water molecule can  form as a function of its distance to the
hydrophobe and hydrophobe radius.  A   boundary hydrogen bond may be slightly altered energetically compared to
the bulk one, but such  alteration is still uncertain$^{24-26}$ and will be neglected hereafter.  

In the probabilistic hydrogen bond (PHB) approach,$^{27}$  a water molecule is considered to have four
arms each capable of forming a single hydrogen bond. The configuration of four hydrogen-bonding (hb)
arms is rigid and symmetric (tetrahedral) with the inter-arm angles  $\alpha=109.47^\circ$ (Fig.1).  Each
hb-arm can  adopt a continuum of orientations  subject to the constraint of tetrahedral rigidity. A
water molecule can form a hydrogen bond with  another molecule only when the tip of any of its hb-arms
coincides with  the second molecule. The length of a hb-arm thus equals the  length of a hydrogen bond
$\eta$, assumed independent of whether the molecules are in bulk or near a 
hydrophobe.   The characteristic  length of pairwise interactions between water molecules
and  molecules constituting the hydrophobe is also assumed to be $\eta$.  

The location of a water molecule is determined by the distance $r$ from its center to  the
center of the hydrophobe which is also chosen as the origin of the spherical coordinate
system. The distance $x$ between water molecule and hydrophobe is defined as $x=r-R$ (Fig.1).

Denote the number of hydrogen bonds per bulk water molecule by $n_b$ and the {\em average}  number of hydrogen
bonds per boundary water molecule by $n_s$. The latter is a  function of radius $R$ and distance $x$, i.e.,
$n_s\equiv n_s(R,x)$. If $x>2\eta$, the number of hydrogen bonds that the water  molecule can form is assumed
to be unaffected by the hydrophobe:  $n_s(R,x)=n_b$ for $x\ge 2\eta$. On the other hand, the function
$n_s(R,x)$  attains its minimum at $x=\eta$,  because at this distance the configurational space available for
neighboring water molecules is most restricted  compared to bulk water. A spherical  layer of thickness $\eta$
from $r=R+\eta$ to  $r=R+2\eta$ is referred to as the solute hydration layer (SHL). 

In the spirit of the PHB approach$^{27}$  let us represent the function $n_s=n_s(R,x)$ as   
\beq n_s=k_1b_1+k_2b_1^2+k_3b_1^3+k_4b_1^4,\eeq 
where $b_1$ is the probability that one of the hb-arms (of a bulk water molecule) can form a 
hydrogen bond and the coefficients  $k_1,k_2,k_3$, and $k_4$ depend on $R$ and $x$, and so does 
$n_s$. Equation (1) assumes that  the {\em intrinsic} hydrogen-bonding 
ability of a water molecule (the tetrahedral configuration of its hb-arms and their lengths and
energies) is unaffected by the hydrophobe. 

The functions $k_1\equiv k_1(R,x),k_2\equiv k_2(R,x),k_3\equiv k_3(R,x)$, and $k_4\equiv k_4(R,x)$
can be  evaluated by using geometric considerations (see the Appendix).  They all become equal to $1$ at $x\ge
2\eta$, where eq.(1) reduces to its bulk analog,  $n_b=b_1+b_1^2+b_1^3+ b_1^4$ (see the Appendix). Since 
experimental data on $n_b$ are readily available, one can
find $b_1$ as a positive solution (satisfying $0<b_1<1$)  of the latter equation. 

Thus, equation (1) provides an efficient pathway to $n_s$ as a function of $x$ and $R$.  It takes into account 
the  constraint that near the hydrophobe some  orientations of  the  hb-arms of a boundary water molecule
cannot lead to the formation of hydrogen bonds. This constraint depends on the  distance  betwe water molecule
and hydrophobe and on the hydrophobe radius, whence the $R$- and $x$-dependence  of $k_1, k_{2}, k_{3},$ and
$k_{4}$. 

Figure 2 presents the function $n_s(R,x)$ for a spherical hydrophobe immersed in water at temperature
$T=293.15$ K, which corresponds to  $n_b=3.65$ hence $b_1=0.963707$. In Fig.2a, $n_s$ is plotted vs
$\xi\equiv (x/\eta-1)$ for various radii $R$. As expected,$n_s$  monotonically increases from its
minimum at $x=\eta$ to its maximum bulk value 
$n_b$ at $x=2\eta$.  For a flat hydrophobic surface ($R=\infty$), molecular dynamics simulations$^{28,29}$ 
previously reported such behavior of $n_s$  (although with some oscillations in ref.29).  In
Fig.2b, $n_s$  is shown as a function of $R$ at different distances $x$. At any $x$, $n_s$
monotonically decreases from its maximum for the smallest particle $R=0$ to its minimum 
for the largest particle ($R=\infty$). Besides, for any fixed $x$, as $R$ 
increases from $0$  to $\infty$, $n_s$ approaches its  asymptotic value for a flat 
hydrophobic surface, $n_s(\infty,x)$, for particles of radii  as small as $R\approx 30\eta$. 

\section{Implementation of the probabilistic hydrogen bond model in the density functional theory}

The fluid density distribution near a rigid surface can be efficiently studied by using 
computer simulations  or DFT.$^{30-32}$  As an illustration of the PHB approach, let us implement it into DFT.
The  latter usually treats  the interaction of fluid molecules with a foreign (impenetrable) substrate
in  the mean-field  approximation whereby  every fluid molecule is considered to be subjected to an
external potential, due to its pairwise interactions with the substrate molecules.$^{31,32}$  The
substrate effect  on the ability of fluid (water) molecules to form hydrogen bonds had been
previously  neglected.  However, using the PHB model,  one can explicitly implement that
effect in the DFT formalism and clarify its role in the length-scale dependence of hydrophobic
hydration. 

To apply DFT to the thermodynamics of hydrophobic phenomena, it is
necessary to know the total  external potential field  $U_{\mbox{\tiny ext}}^{\mbox{\tiny }}\equiv
U_{\mbox{\tiny ext}}^{\mbox{\tiny }}(R,x)$  whereto a water molecule is subjected near 
a hydrophobic particle.  This potential can be written as 
\beq U_{\mbox{\tiny ext}}=U_{\mbox{\tiny ext}}^{\mbox{\tiny p}}+
U_{\mbox{\tiny ext}}^{\mbox{\tiny h}},\eeq
where $U_{\mbox{\tiny  ext}}^{\mbox{\tiny h}}\equiv U_{\mbox{\tiny ext}}^{\mbox{\tiny h}}(R, x)$
is the  water-water hydrogen bond contribution to $U_{\mbox{\tiny ext}}$, and 
$U_{\mbox{\tiny ext}}^{\mbox{\tiny p}}\equiv U_{\mbox{\tiny ext}}^{\mbox{\tiny p}}(R, x)$ 
represents the external {\em pairwise} potential exerted by all the molecules constituting the
hydrophobe on a water molecule. 

While various models were designed$^{31-33}$ for $U_{\mbox{\tiny ext}}^{\mbox{\tiny p}}$, 
the hydrogen bond contribution $U_{\mbox{\tiny ext}}^{\mbox{\tiny h}}$  
had been conventionally neglected until recently$^{34,35}$. 
This contribution, $U_{\mbox{\mbox{\tiny ext}}}^{\mbox{\tiny h}}$,  is due to  the 
deviation of $n_s$ from $n_b$  as well as  the (possible) deviation of $\eps_s$ from $\eps_b$ (the
latter effect is neglected hereafter). It can be determined as 
\beq U_{\mbox{\tiny ext}}^{\mbox{\tiny h}}=\frac1{2}(\eps_sn_s-\eps_bn_b).\eeq 

The first term on the RHS of eq.(3) represents the total energy of hydrogen bonds  of a water molecule
at a distance $x$ from  the surface of a particle of radius $R$,  whereas the second term is the
energy of its hydrogen bonds in bulk (at $x\rightarrow \infty$); the factor $1/2$ is needed to
prevent  double counting the energy because every   hydrogen bond and its energy, either $\eps_s$ or
$\eps_b$, are shared between two  molecules  (in refs.34 and 35 the analogous equation for a planar
surface was mistyped,  as the factor $1/2$ was  missing). Note that $U_{\mbox{\tiny ext}}^{\mbox{\tiny
h}}(R,x)\ne 0$ only for  $\eta\le x \le 2\eta$.

In DFT, the grand thermodynamic potential $\Omega$ of a nonuniform single component
fluid, subjected to an external potential  $U_{\mbox{\tiny ext}}$ (representing the hydrophobe),  
is a functional of the number density $\rho(\bf{r})$ of fluid molecules 
\be \Omega[\rho(\bf{r})]&=&\mathcal{F}_{\mbox{\tiny h}}[\rho({\bf r})] 
+ \frac1{2}\int\int d{\bf r} d{\bf r'}\,\rho({\bf r})\rho({\bf r'}) 
\phi_{\mbox{\tiny at}}(|\bf{r}-\bf{r'}|)\nonumber \\
&+&\int d {\bf r}\, U_{\mbox{\tiny ext}}({R,\bf r})\rho({\bf r})-
\mu\int d{\bf r}\, \rho({\bf r}),\ee
where $\mathcal{F}_{\mbox{\tiny h}}[\rho({\bf r})]$ is the intrinsic Helmholtz free energy
functional of hard sphere fluid, $\mu$ is the
chemical potential, and  $\phi_{\mbox{\tiny at}}(|\bf{r}-\bf{r'}|)$ is the attractive
part of the interaction potential between two fluid molecules located at ${\bf r}$ and ${\bf r'}$;
the integrals are taken over the volume $V$ of the system. Among various models for 
$\mathcal{F}_{\mbox{\tiny h}}[\rho({\bf r})]$, the weighted density approximation (WDA)$^{30,36,37}$ 
with a weight function independent of weighted density represents  
an optimal combination of
accuracy and simplicity. It is non-local with
respect to $\rho({\bf r})$; it takes into account short-ranged correlations and  captures 
the fluid density oscillations near a hard wall. We hereafter adopt the WDA version of DFT.

The key element of WDA  is the weighted density $\widetilde{\rho}({\bf r})$ 
determined in terms of $\rho({\bf r})$ via an implicit equation 
\beq \widetilde{\rho}({\bf r})=\int d{\bf r'}\,\rho({\bf r'})
w(|{\bf r'}-{\bf r}|;\widetilde{\rho}({\bf r})),\eeq  
where $w(|{\bf r'}-{\bf
r}|;\widetilde{\rho}({\bf r}))$ is the weight function. Although in more sophisticated
versions$^{30,36}$ of WDA $w(|{\bf r'}-{\bf r}|;\widetilde{\rho}({\bf r}))$ depends on
$\widetilde{\rho}({\bf r})$, we will hereafter adopt its simpler version wherein  the weight
function is independent$^{30,37}$ of $\widetilde{\rho}({\bf r})$.   

For hydrophobic hydration in an open system of constant  $\mu$, $V$, and $T$ 
(grand canonical ensemble), 
the  equilibrium density profile is obtained
by minimizing  $\Omega[\rho(\bf{r})]$ with respect to $\rho({\bf r})$.  The
corresponding Euler-Lagrange equation can be written as 
\be \mu=k_BT\ln(\Lambda^3\rho({\bf r}))+W({\bf r};\rho({\bf r})),\ee 
where $\Lambda=(h^2/2\pi mk_BT)^{1/2}$ is the thermal de Broglie wavelength of a molecule  of mass
$m$ ($h$ and $k_B$ being Planck's and Boltzmann's constants)  and 
$W({\bf r};\rho({\bf r}))$ is a function of ${\bf r}$ and a functional of $\rho({\bf r})$:$^{28-30}$
\be W({\bf r};\rho({\bf r}))&=&U_{\mbox{\tiny ext}}({\bf r})+
\int d{\bf r'} \,\rho({\bf r'}) \phi_{\mbox{\tiny a}}(|{\bf r}-{\bf r'}|)\nonumber\\
&+&
\Delta \psi_{\mbox{\tiny h}}(\widetilde{\rho}({\bf r}))+
\int d{\bf r'}\,\rho({\bf r'})\,\Delta\psi'_{\mbox{\tiny h}}(\widetilde{\rho}({\bf r'}))
w(|{\bf r'}-{\bf r}|).\ee 
Here $\phi_{\mbox{\tiny a}}(|{\bf r}-\bf{r'}|)$ is the
attractive part of the interaction potential between two fluid molecules located at ${\bf r}$
and ${\bf r'}$, whereas $ \Delta \psi_{\mbox{\tiny h}}(\rho)$ 
is the configurational part of the free energy of hard sphere fluid per molecule, with 
$\Delta\psi'_{\mbox{\tiny h}}(\rho)\equiv d\Delta\psi_{\mbox{\tiny h}}(\rho)/d\rho$.

The hydrophobe being spherical, the external potential is a function of a
single variable $x=r-R$, and the equilibrium density profile obtained from eq.(6) is a function of
a single variable $r$: $\rho({\bf r})=\rho(r)$. The substitution of $\rho(r)$ into eq.(4)
provides the grand thermodynamic potential $\Omega$ of the non-uniform fluid with a hydrophobe 
therein.  The grand canonical free energy of hydration is 
$\De \Omega_{\mbox{\tiny }}=\Omega-\Omega_0$, 
where $\Omega_0$ is the grand thermodynamic potential  of uniform liquid water 
without a hydrophobe therein.

\section{Numerical Calculations}

For a numerical illustration, we considered the hydration of a spherical hydrophobe 
(taking  $R/\eta=1,3,5,7,10,15,20,30,50,100$) in the model water  at $T=293.15$ K and 
$\mu=-11.5989$ $k_BT$ corresponding to its two-phase equilibrium. The liquid state of bulk water
was ensured by imposing the appropriate boundary condition onto eq.(6), $\rho(x)\rightarrow \rho_l$ as
$x\rightarrow \infty$, with $\rho_l$ the bulk liquid density.  The densities  $\rho_v$ and
$\rho_l$ of coexisting vapor and liquid, respectively, are determined  by solving the equations 
$\left.\mu(\rho,T)\right|_{\rho=\rho_v}=\left.\mu(\rho,T)\right|_{\rho=\rho_l},\;\;\;\;\;\;\;\;
\left.p(\rho,T)\right|_{\rho=\rho_v}=\left.p(\rho,T)\right|_{\rho=\rho_l}$, requiring 
the chemical potential $\mu\equiv\mu(\rho,T)$ and pressure  $p\equiv 
p(\rho,T)$ to be the same throughout both coexisting phases. 

The chemical potential of a uniform hard sphere fluid  $\mu_{\mbox{\tiny h}}$  and the 
configurational part 
$\Delta\psi_{\mbox{\tiny h}}\equiv\Delta\psi_{\mbox{\tiny h}}(\rho,T)$ of the
free energy of a hard sphere fluid were modeled in 
the Carnahan-Starling approximation,$^{31,32,38}$ 
whereas for the weight function $w(|{\bf r'}-{\bf r}|;\widetilde{\rho}({\bf r}))$ in eqs.(5),(7)  
we adopted a 
$\widetilde{\rho}$-independent version$^{37}$ 
$$\Delta \psi_{\mbox{\tiny h}}=k_BT\frac{\xi\,(4-3\xi)}{(1-\xi)^2},\;\;\;\;\; 
w(r_{12})=\frac3{\pi\eta^4}(\eta-r_{12})\Theta(\eta-r_{12}),$$ 
with $\Theta(u)$ being the Heaviside (unit-step) function. 

The pairwise  interactions of water
molecules were modeled by using the Lennard-Jones (LJ) potential with the energy parameter 
$\eps_{\mbox{\tiny ww}}=3.79\times 10^{-14}$ erg  and the diameter $d$ of a model  molecule  set
to be $\eta$.  The attractive part $\phi_{\mbox{\tiny at}}$
of pairwise water-water interactions was modeled via the Weeks-Chandler-Anderson perturbation
scheme.$^{39}$ 
The interaction potential between water molecule and molecule of a 
hydrophobe was assumed to be of LJ type with an energy parameter $\eps_{\mbox{\tiny
wp}}$ and a length parameter $\eta$.  Integrating this interaction with respect to the position of
the molecule of the hydrophobe over the hydrophobe volume $V_R=4\pi R^3/3$, one
can obtain  the pairwise contribution $U_{\mbox{\tiny ext}}^{\mbox{\tiny p}}$ into $U_{\mbox{\tiny
ext}}$.  We assumed the dimensionless number density of molecules in the hydrophobe to
be $\rho_p\eta^3\approx 1$ and considered five values for $\eps_{\mbox{\tiny
wp}}/\eps_{\mbox{\tiny ww}}$  (0.6, 0.66, 0.75, 0.9, 1.2)  to mimic various degrees
of hydrophobicity. 
The density profiles and free energies of hydration thus obtained are shown in
Figures 3-5.

Figure 3 presents the density profiles near a spherical hydrophobe of  radius $R$ for
$R/\eta=1,3,5,7,10,15,30,100$ and $\eps_{\mbox{\tiny wp}}/\eps_{\mbox{\tiny ww}}=0.75$ .  The profiles in this
Figure (as well as in Figure 4) are
presented in a coordinate system with the origin $r=0$ located in the  center of the hydrophobe; there are no
water  molecules at $r<R+\eta$ (the space $0<r\le R$ is occupied by the hydrophobic sphere and  the layer
$R<r<R+\eta$ is excluded to fluid molecules).As clear, the hydrophobe radius greatly affects the distribution of
vicinal water molecules.  The oscillations in the density profile gradually disappear as  $R$  increases. They 
are well pronounced for $R/\eta=1$, but virtually  non-existent for particles $R/\eta\ge 7$.  As $R$ increases,
a thin depletion layer around the particle  (virtually non-existent for $R/\eta=1$) becomes more  developed,
with its density approaching that of vapor and its thickness approaching $\eta$. 

This is consistent with the largely accepted wisdom concerning the much discussed issue whether or not there is
a vapor-like layer near a large hydrophobe in liquid water. As now widely agreed upon, even if (and when) such a
layer exists, it should be expected to be of molecular thickness only.$^{10,11,19,21-23}$ 

Furthermore, the behavior of fluid density profiles in Fig.3 is 
consistent with our previous finding$^{34,35}$ that  the hydrogen bond contribution to the
external potential plays a crucial role in the formation of a thin ``strong depletion" layer 
(of density much
lower than liquid and  of thickness of a molecular diameter) between liquid water and 
{\em planar} hydrophobic surface even for weakly
hydrophobic surfaces (with high $\eps_{\mbox{\tiny wp}}/\eps_{\mbox{\tiny ww}}$).  Indeed, as $R$
increases, the geometric constraint on the ability of a vicinal water molecule to form hydrogen
bonds strengthens, the repulsive contribution  
$U_{\mbox{\tiny ext}}^{\mbox{\tiny h}}$ to $U_{\mbox{\tiny ext}}$ increases, whence the 
widening and stronger depletion of the vicinal water layer around the hydrophobe if the ``pairwise"
hydrophobe-water (attractive) contribution $U_{\mbox{\tiny ext}}^{\mbox{\tiny p}}$ 
(determined by the ratio $\eps_{\mbox{\tiny wp}}/\eps_{\mbox{\tiny ww}}$) is not too large (by absolute value). 

In order to further clarify the effects of water-hydrophobe attraction and hydrophobe radius  on fluid (water)
density profiles, they are plotted in Figure 4 for three different values of  $\eps_{\mbox{\tiny
wp}}/\eps_{\mbox{\tiny ww}}$ and two different radii $R$, namely, $R/\eta=1$ (Fig.4a) and $R/\eta=15$  (Fig.4b).
Three profiles shown in each of Figs.4a and 4b correspond to $\eps_{\mbox{\tiny wp}}/\eps_{\mbox{\tiny
ww}}=0.6,0.75,1.2$ (from bottom to top, respectively).  As clear, the strengthening of pairwise intermolecular
fluid-hydrophobe interactions by 100\% has a little effect on density profiles near a sufficiently large
hydrophobe ($R/\eta=15$); the thickness of the depletion layer remains virtually unaffected (roughly equal to
$\eta$) and the fluid density therein remains several orders of magnitude lower than its bulk value (i.e., the
depletion layer remains vapor-like). On the other hand, for a molecular size hydrophobe ($R/\eta=1$), the
increase of $\eps_{\mbox{\tiny wp}}/\eps_{\mbox{\tiny ww}}$  by 100\% leads to a drastic change in the nature of
the water depletion layer near the hydrophobe; it becomes significantly narrower and from being a vapor-like one
transforms into a liquid-like one. These results are in qualitative agreement with the previously reported ones
obtained via molecular dynamics simulations$^{40}$ of the SPCE water model
and via Monte Carlo simulations$^{41}$  of the TIP4P water model. The latter study also reported
the analogous behavior  for all fluids, including nonassociating ones (without hydrogen-bonding ability). That
is somewhat dissimilar from our previous studies$^{34}$ showing  that that even for a relatively strong
hydrophobic {\em planar} surface (with low  $\eps_{\mbox{\tiny wp}}/\eps_{\mbox{\tiny ww}}$)  the conventional
contribution to the external potential (due to pairwise interactions between a water molecule and those of the
substrate) cannot cause the formation of a vapor-like layer near the surface, although it does lead to the
formation of a depletion layer with  a weak decrease in the vicinal fluid density compared to the bulk one. 

In the framework of the proposed approach,  the thickness of and the density in the depletion layer are
determined by the interplay of two  effects.  On one hand, the constraint on the configurational space available
to water molecules, wherewith a selected molecule can form hydrogen bonds, is taken account of via the function
$n_s(R,x)$, eq.(1).  On the other hand,  the model incorporates attractive interactions between water
molecule and hydrophobe, eq.(2),  whereof the strength can be characterised by the positive parameter
$\eps_{\mbox{\tiny wp}}$. For a given thermodynamic state of the system and the nature of the hydrophobe 
(represented by
$\eps_{\mbox{\tiny wp}}$),  the result of this interplay  naturally depends on the hydrophobe size (radius). When the
former effect predominates over the latter, the depletion layer is vapor-like. Otherwise, for relatively weakly 
hydrophobic particles (that are not too large), the vicinal water layer is just slightly depleted 
compared to the bulk liquid.

Figure 5a presents the grand canonical free energy of hydrophobic  hydration  $\De \Omega_{\mbox{\tiny }}$ as
a function of the hydrophobe radius $R$ (note that the curves are provided only for guiding the eye; the actual
calculated points are at $R/\eta=1,3,5,7,10,15,20,30,50$, and $100$).  The intrinsic hydrophobicity of the
particles is assumed to be independent  of $R$, with  $\eps_{\mbox{\tiny wp}}/\eps_{\mbox{\tiny ww}}=0.75$.
The hydration free energy is expressed in  units of $k_BT$ per ``dimensionless unit area"; the  dimensionless
$\overline{\De\Omega}$ in Figure 4 is  obtained by dividing $\De \Omega$  by $k_BT$ and by $4\pi R^2/\eta^2$. 
The variable  sensitivity of $\overline{\De\Omega}$ to $R$ is a  clear indication that the hydration of small
and large length-scale particles occurs via different  mechanisms and that the hydrogen bond contribution to
$U_{\mbox{\tiny ext}}$  plays a key role in this process. The model predictions for $\overline{\De\Omega}$ for
small $R$'s are consistent  with the  experimental data on the hydration free energy of methane, ethane, 
propane, and $n$-butane at  the temperature $T=300$ K, as compiled in ref.8;   considering a  methane molecule
as a sphere and ethane, propane, and n-butane molecules as cylinders, one can roughly estimate the
experimental $\overline{\De\Omega}_{\mbox{\tiny }}$ to be $0.6$ for methane, $0.4$ for ethane, and $0.3$ for
propane and $n$-butane.  

The dominant role of the hydrogen bond network in hydrophobic hydration is emphasized in  Figure 5b where
$\overline{\De\Omega}$ is plotted vs  $\eps_{\mbox{\tiny wp}}/\eps_{\mbox{\tiny ww}}$  for various radii $R$.
Each curve in Fig.5b corresponds to a fixed $R$, with $R/\eta=1,3,7,15,30$  from bottom to top (again the
curves are provided only for guiding the eye; the actual calculated points are at  $\eps_{\mbox{\tiny
wp}}/\eps_{\mbox{\tiny ww}}=0.6,0.66,0.75,0.9$, and $1.2$). As expected,  the hydration free energy  per unit
area  decreases with increasing degree of hydrophobicity and, for small enough particles, 
$\overline{\De\Omega}$ may even become negative.  The model predictions suggest that the hydration of even
{\em apolar } particles  of radii $R\lesssim 3\eta$ may be thermodynamically favorable (the hydration free
energy being negative)  if the pairwise (LJ-type) interactions between a water molecule and a molecule
constituting the hydrophobe are comparable with or stronger than the (LJ-type) interactions between two water
molecules, i.e., if $\eps_{\mbox{\tiny wp}}/\eps_{\mbox{\tiny ww}}\gtrsim 1$. This result clarifies 
some previous simulational and theoretical observations$^{18-20}$ that two inert gas molecules would prefer to 
form a solvent-separated pair rather than a contact pair (dimer).

\section{Conclusions}

Concluding, we emphasize that the PHB approach allows one to obtain an analytic expression for the average
number of hydrogen bonds per water molecule near a spherical hydrophobic particle as a function of both the
particle radius and the distance between water  molecule and hydrophobe surface. This function can serve as a
foundation  for elucidating various aspects of hydrophobic  phenomena, particularly their length-scale
dependence,  either via computer simulations (Monte Carlo and Molecular Dynamics) or  DFT.  For example,  this
function allows one to explicitly identify  an  additional contribution to the external  potential exerted by
the hydrophobe on a water molecule; it is due to the alteration of water hydrogen bonding near a hydrophobe. 
Thus, one can efficiently  implement  the hydrogen bonding ability of water molecules  in DFT to examine the
particle size  dependence of hydrophobic hydration. 

As a numerical illustration of the combined PHB/DFT approach, we have studied the hydration of spherical
particles of various radii and various hydrophobicity in a model water. The numerical results for the hydration
free energy of small size hydrophobes  are consistent with the experimental data on the hydration of small
alkanes. The free energy of hydration per unit area of a spherical particles is predicted to have a varying
sensitivity to the particle radius which is a clear indication that the hydration of small and large
length-scale particles occurs via different  mechanisms. On the other hand, the model predictions suggest that
the hydration of even {\em apolar } particles  of small enough radii may be thermodynamically favorable if the
pairwise (dispersion)  attraction between a water molecule and a molecule  constituting the hydrophobe is
comparable with or stronger than the pairwise (dispersion) attractions  between two water molecules. This
result at least partially clarifies  some previous simulational and theoretical observations  (rather
counterintuitive from the conventional point of view on hydrophobicity) that two inert gas molecules would
prefer to  form a solvent-separated pair rather than a contact pair (dimer).

Note that in the PHB approach the tetrahedral rigidity (both geometric and energetic)  of the hb-arms
of a water molecule  is assumed only for the analytical simplicity. It can be eliminated to allow for 
the geometric deformation and energetic alteration of this configuration depending on the sequence in
which the hb-arms are engaged or due to the proximity of the water molecule to the hydrophobe. 
These modifications will just render the model more complicated for analytical treatment and can be
expected to relatively weakly affect the model predictions. 

\newpage
\appendix
\section*{Appendix. Derivation of the coefficients $k_1,k_2,k_3$, and $k_4$ as functions of $R$ and $x$}
\newcounter{s2e}
\newcommand{\ale}{\setcounter{s2e}{\value{equation}}%
\stepcounter{s2e}\setcounter{equation}{0}\renewcommand{\theequation}{A\arabic{equation}}}
\ale 

To find the function $n_s(R,x)$, first consider its bulk analog, $n_b$,  and represent   it
as (see ref.[27] in the main text) 
\beq n_b=b_1+b_{2(1)}+b_{3(2,1)}+b_{4(3,2,1)},\eeq  
where $b_1$ is the probability that
one of the hb-arms (of a bulk water molecule) can form a  hydrogen bond, $b_{2(1)}$ is the 
probability that a second hb-arm can form a hydrogen bond subject  to the condition that one of
the hb-arms has already formed a bond, $b_{3(2,1)}$ is the probability that a third hb-arm can
form  a hydrogen bond subject to the condition that two of the hb-arms have already formed bonds,
and $b_{4(3,2,1)}$ is the probability that the fourth hb-arm can form a bond subject to the
condition that three of the  hb-arms have already formed bonds. 

Note that the probability $b_1$  can  be formally represented as a product $b_1=P_{S\rightarrow
N}P_{N\rightarrow S}$, where   $P_{S\rightarrow N}$ is the probability that the tip of any hb-arm
of  molecule $S$ roughly coincides with molecule $N$ and $P_{N\rightarrow S}$ is  the probability
that the tip of any  hb-arm of molecule $N$ roughly coincides with molecule $S$.  (Similar
considerations are valid for $b_{2(1)},\;b_{3(2,1)}$, and $b_{4(3,2,1)}$ as well).  Neither
$P_{S\rightarrow N}$  nor $P_{S\rightarrow S}$ can be found in the framework of our simple model,
but their product (i.e., $b_1$) can be determined from readily available experimental and
simulational data on $n_b$. 

Indeed, in the chosen model of a water molecule the events of formation of bonds by the hb-arms
(in bulk water) can be considered as independent of each other, so that 
$b_{2(1)}=b_1^2,\;\;b_{3(2,1)}=b_1^3,\;\;b_{4(3,2,1)}=b_1^4.$ Thus, the probability $b_1$ can be
evaluated as the positive solution of the equation $n_b=b_1+b_1^2+b_1^3+ b_1^4$ satisfying the
condition $0<b_1<1$. The latter representation of $n_b$ implies that  the {\em intrinsic}
hydrogen bonding ability of each arm is independent of whether the other arms  have been already
engaged in hydrogen bonds or not.  That is, when the first hb-arm of a water molecule forms an
actual bond, the electron  density distribution in a water molecule determining the ability of
the  other three hb-arms (that are not engaged yet) to form bonds  (and their potential
orientations) remains unaffected  (there is no issue with the availability of water molecules
necessary for the selected bulk molecule to form bonds). 

For a boundary water molecule, let us represent $n_s$ in a form: 
\beq n_s=s_1+s_{2(1)}+s_{3(2,1)}+s_{4(3,2,1)}.\eeq 
Here $s_1\equiv s_1(R,x),\; s_{2(1)}\equiv s_{2(1)}(R,x),\; 
s_{3(2,1)}\equiv s_{3(2,1)}(R,x),\; s_{4(3,2,1)}\equiv s_{4(3,2,1)}(R,x)$ are 
probabilities analogous to  $b_1, b_{2(1)}, b_{3(2,1)}, b_{4(3,2,1)}$ subject to the  constraint
that some  orientations of the  hb-arms cannot lead to the formation of hydrogen bonds because of
the proximity to the hydrophobic  particle. The severity of this constraint depends on the
distance of the water molecule to the particle, hence the $x'$-dependence of $s_1, s_{2(1)}, s_{3(2,1)},
s_{4(3,2,1)}$.  Again, as a first approximation  the intrinsic hydrogen-bonding ability of a water molecule 
(i.e., the tetrahedral configuration of its hb-arms and their lengths and energies) can be
considered to be unaffected by its proximity to the hydrophobic particle so that the  latter only
restricts the configurational space available to  other water molecules  necessary for this 
boundary water molecule to
form hydrogen bonds.   Thus, one can relate  $s_1, s_{2(1)}, s_{3(2,1)}$,  and $s_{4(3,2,1)}$ to 
$b_1,\;\;b_{2(1)},\;\;b_{3(2,1)}$, and $b_{4(3,2,1)}$, respectively, as 
\beq s_1=k_1b_1,\;\;\;s_{2(1)}=k_2b_{2(1)},\;\;\;
s_{3(2,1)}=k_3b_{3(2,1)},\;\;\; s_{4(3,2,1)}=k_4b_{4(3,2,1)},\eeq
where the coefficients $k_1,\;k_2,\;k_3$, and $k_4$ are functions of $R$ and $x$ (with their dependence on
the boundary water molecule orientations averaged) and can be evaluated by using geometric considerations.

\subsubsection*{Coefficient $k_1$}
The coefficient $k_1$ is calculated by taking into account that a boundary water molecule 
can form a
hydrogen bond ``almost" like a bulk molecule except for the constraint that 
the tip of the  hb-arm (arm 1) must not be too close to the surface of the hydrophobic particle of
radius $R$. 
Select an arbitrary water molecule at a distance $\eta \le x\le 2\eta$ from that surface    
and denote it $S$ (Figure 4). Any  of its hb-arms can form a hydrogen 
bond if the tip
of the arm is located anywhere on a sphere of radius $\eta$ (centered at $S$) 
from which a spherical
cap is cut out by the sphere $ib$ of radius $R+\eta$, inner boundary of the 
SHL of particle $R$. Denoting the corresponding solid angle by 
$\Omega_s(R,x)$, one can write  
\beq \Omega_s(R,x)=2\pi \int_{0}^{\Theta_{1x}^{M}(x)}d\Theta_1\,\sin(\Theta_1),\eeq
where $\Theta_1$ is the angle between hb-arm 1 and radial axis $r$ (with the origin in the center
of the hydrophobe and passing through the molecule $S$), with 
$\Theta_{1x}^{M}(R,x)\equiv \arccos[-(2(R+\eta+x)+\eta^2+x^2)/2(R+\eta+x)\eta]$ is the maximum 
angle $\Theta_1$ at which hb-arm 1 can still form a bond. 
The probability $s_1(R,x)$ 
that any one of hb-arms of molecule $S$ can form a hydrogen bond 
is related to $b_1$ via 
\beq s_1(R,x)=\frac{\Omega_s(R,x)}{\Omega_b}b_1,\eeq
where $\Omega_b=4\pi$. 
Integrating the RHS of eq.(S4), substituting the result into eq.(S5), 
and taking into account eq.(S3), 
one obtains the coefficient $k_1\equiv k_1(R,x)$ to be :
\beq k_1=\frac1{2}\left(1+\frac{(2(R+\eta)x+\eta^2+x^2)}{2(R+\eta+x)\eta}\right).\eeq

\subsubsection*{Coefficient $k_2$}

The coefficient $k_2$ is calculated by
assuming that hb-arm 1 has already formed a bond in an arbitrary orientation $\Theta_1$ (with
respect to the radial axis $r$) and by taking
into account that hb-arm 2 can form a bond subject to the condition that its tip must not 
lie closer to the particle surface than the SHL inner boundary $ib$ of hydrophobe $R$. 

For a bulk water molecule the probability $b_{2(1)}$,  that the second hydrogen bond forms once
the first one has formed is proportional to the full length $L_b$ of the circle $C_2(1)$ formed
by  the possible loci  of the tip of the {\em engaged} second hb-arm (hb-arm 2) subject to the 
restriction that the angle between the two hb-arms remains $\alpha$.  Since the radius of that
circle is equal to $R_{\mbox{\tiny{C}}_{2(1)}}=\eta\sin(\alpha)$, we have 
\beq L_b=2\pi\eta\sin(\alpha).\eeq 

However, for the molecule $S$ the possible loci of the tip of {\em engaged} hb-arm 2  (subject to
the restriction that  the angle between it and hb-arm 1 is $\alpha$) constitute just a part of the
circle of radius $\eta\sin(\alpha)$, - the other part (a circular arc) being excluded by the
proximity of the SHL inner boundary $ib$. The length of the ``{\em available}" part of this circle
is a function of $R,\, x$, and $\Theta_1$ and will be denoted by 
$L_s\equiv L_s(R,x,\Theta_1)$ (by definition, $L_s(R,x,\Theta_1)=L_b$ for $x\ge 2\eta$). 
The probability $s'_{2(1)}$ that molecule $S$ engages in a second hydrogen bond once its hb-arm 
1 has already formed a bond (at given $R,x,\Theta_1$) is proportional to $L_s$: 
\beq s'_{2(1)}\equiv s'_{2(1)}(R,x,\Theta_1)=\frac{L_s(R,x,\Theta_1)}{L_b}b_{2(1)}\eeq 

Let us introduce the Cartesian coordinate system with the origin $O$  in
the center of particle $R$, axis $z$ coinciding with the radial axis $r$, and axis $y$ directed
from the origin towards the projection of the tip of hb-arm $1$ onto the $x-y$ plane. Depending on
the angle $\Theta_1$, distance $x$, and radius $R$, the circle $C_2(1)$ may either intersect the
SHL inner boundary (which is a sphere of radius $R+\eta$ centered at $0$, hereafter denoted
$S_{SHL}^{ib}$) or not. In the former
case, there may be either two intersection points which can degenerate into one point in the
limiting at some particular orientation for given $R$ and $x$. Assuming that there are two
intersection points of sphere $S_{SHL}^{ib}$ and circle $C_2(1)$, let us denote their Cartesian
coordinates by $x_-,y_-,z_-$ and $x_+,y_+,z_+$. Clearly, $y_{\pm}\equiv y_-=y_+ $, 
$z_{\pm}\equiv z_-=z_+$, and $x_-=-x_+$; one can choose the notation so that $x_+>0$; clearly the
coordinates of both points are functions of $R,x,\Theta_1$. 
 
Further, let us define the angle $\Theta_{10}(R,x)$ to be the value of the angle $\Theta_1$ 
at which the circle $C_2(1)$ just ``touches" the sphere $S_{SHL}^{ib}$, and introduce 
$x^0\equiv x^0(R)$ and $R^0$ as 
the solutions of equations $\Theta_{10}(R,x^0)=0$ (with respect to $x^0$) and 
$x^0(R^0)=0$ (with respect to $R^0$), respectively. One can thus obtain  $R^0\approx 0.50009$,  
\beq
x^0=-R-\eta\cos(\alpha)+\sqrt{R^2+2R\eta+\eta^2\cos^2(\alpha)},\;\;\;
\Theta_{10}(R,x)=-\alpha+\Theta_{1x}^{M}(R,x), \eeq 

For $0\le R <R^0$, one can show that 
\beq L_s=\left\{ \begin{array}{ll} 
L_b &\;\;\;\;\;\; \mbox{if}\;\;\;\;\;\;   
\Theta_1\in \;\;  [0;\Theta_{10}(R,x)]\\ 
L_{c}^{\pm} & \;\;\;\;\;\;\mbox{if}\;\;\;\;\;\;   
\Theta_1\in\;\;\; ]\Theta_{10}(R,x);\Theta_{1}^{\mbox{\tiny{mM}}}(R,x)]\\
L_b &\;\;\;\;\;\; \mbox{if} \;\;\;\;\;\;  
\Theta_1\in \;\;\;  ]\Theta_{1}^{\mbox{\tiny{mM}}}(R,x);\Theta_{1x}^{M}(R,x)]
\end{array} \right.\eeq 
where 
$\Theta_{1x}^{\mbox{\tiny{mM}}}(R,x)=\min[\Theta_{1x}^{m}(R,x),\Theta_{1x}^{M}(R,x)]$,  
$\Theta_{1x}^{m}(R,x)=2\pi-\alpha-\Theta_{1x}^{M}(R,x)$, and 
\beq  L_c^{\pm}=\left\{\begin{array}{ll} 2\phi_{\mbox{\tiny{C}}}^{\pm}R_{\mbox{\tiny{C}}_{2(1)}} &\;\;\;\;\;\; \mbox{if}\;\;\;\;\;\;   
y_{\pm}\ge y_{\mbox{\tiny{C}}_{2(1)}}^o,\\ 
2(\pi-\phi_{\mbox{\tiny{C}}}^{\pm})R_{\mbox{\tiny{C}}_{2(1)}} &\;\;\;\;\;\; \mbox{if}\;\;\;\;\;\;   
y_{\pm}< y_{\mbox{\tiny{C}}_{2(1)}}^o,
\end{array} \right.  \eeq 
with $\phi_{\mbox{\tiny{C}}}^{\pm}=\arcsin(x_{+}/\eta\sin(\alpha))$ and 
$y_{\mbox{\tiny{C}}_{2(1)}}^o$ the $y$-coordinate of the center of the circle $C_2(1)$. 
The expression for  $L_c^{\pm}$ can be also written as 
\beq  L_c^{\pm}=
\left\{
\begin{array}{ll} 2\phi_{\mbox{\tiny{C}}}^{\pm}R_{\mbox{\tiny{C}}_{2(1)}} 
&\;\;\;\;\;\; \mbox{if}\;\;\;\;\;\; 
\Theta_1\in\;\;\; [\Theta_{1n}(R,x);\Theta_{1}^{\mbox{\tiny{$\phi$}}}(R,x)],\\ 
2(\pi-\phi_{\mbox{\tiny{C}}}^{\pm})R_{\mbox{\tiny{C}}_{2(1)}} &\;\;\;\;\;\; 
\mbox{if}\;\;\;\;\;\; \Theta_1\in\;\;\; ]\Theta_{1}^{\mbox{\tiny{$\phi$}}}(R,x);
\Theta_{1x}^{\mbox{\tiny{mM}}}(R,x)],
\end{array} 
\right.  
\eeq 
where $\Theta_{1}^{\mbox{\tiny{$\phi$}}}(R,x)$  is the value of the angle $\Theta_1$ at which
$x_{+}=R_{\mbox{\tiny{C}}_{2(1)}}$. 

Considering $R^0\le R<\infty$ 
one can show that for $x\in[1;x^0(R)]$ 
\beq L_s=\left\{ \begin{array}{ll} 
0 &\;\;\;\;\;\; \mbox{if}\;\;\;\;\;\;   
\Theta_1\in \;\;  [0;\Theta_{1n}^{\mbox{\tiny{}}}(R,x)],\\ 
L_{c}^{\pm} & \;\;\;\;\;\;\mbox{if}\;\;\;\;\;\;   
\Theta_1\in\;\;\; ]\Theta_{1n}^{}(R,x);\Theta_{1x}^{\mbox{\tiny{mM}}}(R,x)],\\
L_b &\;\;\;\;\;\; \mbox{if} \;\;\;\;\;\;  
\Theta_1\in \;\;\;  ]\Theta_{1x}^{\mbox{\tiny{mM}}}(R,x);\Theta_{1x}^{M}(R,x)],
\end{array} \right.\eeq 
with $\Theta_{1n}^{}(R,x)=\alpha - \Theta_{1x}^{M}(R,x)$, 
whereas for $x\in[x^0(R);2\eta]$ 
\beq L_s=\left\{ \begin{array}{ll} 
L_b &\;\;\;\;\;\; \mbox{if}\;\;\;\;\;\;   
\Theta_1\in \;\;  [0;\Theta_{10}(R,x)],\\ 
L_{c}^{\pm} & \;\;\;\;\;\;\mbox{if}\;\;\;\;\;\;   
\Theta_1\in\;\;\; ]\Theta_{10}(R,x);\Theta_{1x}^{\mbox{\tiny{mM}}}(R,x)],\\
L_b &\;\;\;\;\;\; \mbox{if} \;\;\;\;\;\;  
\Theta_1\in \;\;\;  ]\Theta_{1x}^{\mbox{\tiny{mM}}}(R,x);\Theta_{1x}^{M}(R,x)].
\end{array} \right.\eeq 

The probability $s_{2(1)}\equiv s_{2(1)}(R,x)$ defined by eq.(S2), can be obtained by integrating
$s'_{2(1)}(R,x,\Theta_1)$ with respect to the angle $\Theta_1$ 
subject to the constraint that hb-arm 1 has already formed a bond (the distribution of
possible orientations of hb-arm 1 assumed to be uniform):
\beq s_{2(1)}(R,x)=\frac1{\Theta_{1x}^{M}(R,x)}
\int_{0}^{\Theta_{1x}^{M}(R,x)}\mathrm{d}\Theta_1\;s'_{2(1)}(R,x,\Theta_1),\eeq
Therefore, according to eqs.(S3) and (S8), we have 
\beq k_{2(1)}(R,x)=\frac1{\Theta_{1x}^{M}(R,x)}
\int_{0}^{\Theta_{1x}^{M}(R,x)}\mathrm{d}\Theta_1\;\frac{L_s(R,x,\Theta_1)}{2\pi\eta\sin(\alpha)}.
\eeq 

\subsubsection*{Coefficient $k_3$} 

The coefficient $k_3$ is found by assuming that hb-arms 1 and 2 of molecule $S$ (see Figure
1)  have already formed bonds in arbitrary orientations, determined by angles $\Theta_1$ and
$\Theta_2$ that they form with the $x'$ axis and  calculating the probability that the third
hb-arm (say, hb-arm 3) can also form a bond    subject to the constraint that the angle
between any two hb-arms is equal to $\alpha$.   Certainly, for  hb-arm 3 to form a bond its
tip must not lie to the left of the plane $Ll$ (left boundary of  the SHL).  
The  orientations of hb-arms 1 and 2 are eventually averaged assuming their uniform distributions.

Clearly,  hb-arm 3 can form a bond with the same probability as in the bulk if the location of
molecule $S$ and orientation of its arms 1 and 2 (i.e., $x,\Theta_1,$ and $\Theta_2$)  are
such that the tip of arm 3 is {\em not} located within the inner boundary of the particle SHL. 
Otherwise, hb-arm 3
cannot form a bond at all.  Therefore,  
$s'_{3(2,1)}\equiv s'_{3(2,1)}(R,x,\Theta_1,\Theta_2)$, one can thus write 
\beq s'_{3(2,1)}=\left\{\begin{array}{ll}  b_{3(2,1)} & \;\;\; 
\mbox{if}\;\;\Theta_1\in\;\;[\Theta_{1n}^{(3)}(R,x);\Theta_{1x}^{(3)}(R,x)]\;\;\mbox{and}\;\;
\Theta_2\in\;\;[\Theta_{2n}^{(3)}(R,x,\Theta_1);\Theta_{2x}^{(3)}(R,x,\Theta_1)],\\
0 & \;\;\; \mbox{otherwise},\end{array} \right. \eeq  
where $\Theta_{1n}^{(3)}(R,x)$ and $\Theta_{1x}^{(3)}(R,x)$ 
are the minimum and maximum angles  (both in the range from 0 to $\pi$) 
that hb-arm 1 can have with the $r$-axis for three hydrogen
bonds to form (for given $R$ and $x$);  $\Theta^{(3)}_{2x}(R,x,\Theta_1)$ and 
$\Theta_{2x}^{(3)}(R,x,\Theta_1)$ are the minimum and maximum angles 
(both in the range from 0 to $\pi$) that hb-arm 2 can have with the $t$-axis for three 
hydrogen bonds to form simultaneously (for given $R,x$, and 
$\Theta_{1n}^{(3)}(R,x)\le\Theta_1\le\Theta_{1x}^{(3)}(R,x)$ ).

The mean  probability $s_{3(2,1)}$ that molecule $S$ for given $R$ and $x$ 
forms a third hydrogen bond once its hb-arms 1
and 2 have already formed bonds can be obtained by  averaging $s'_{3(2,1)}$ over all  
possible orientations of 
its hb-arms 1 and 2, i.e., over $\Theta_1$ and $\Theta_2$ (both angles 
distributed uniformly) 
\beq s_{3(2,1)}=\int\limits_{\Theta_{1n}^{(3)}(R,x)}^{\Theta^{(3)}_{1x}(R,x)}
\frac{\mathrm{d}\Theta_1}{\Theta_{1x}^{M}(R,x)-\widetilde{\Theta}_{1n}^{}(R,x)} 
\int\limits_{\Theta_{2n}^{(3)}(R,x,\Theta_1)}^{\Theta_{2x}^{(3)}(R,x,\Theta_1)}
\frac{\mathrm{d}\Theta_2}{\Theta_{2x}^{}(R,x,\Theta_1)-\Theta_{2n}(R,x,\Theta_1)}
s'_{3(2,1)}(R,x,\Theta_1,\Theta_2).\eeq
Here $\widetilde{\Theta}_{1n}^{}(R,x)\equiv H(x^0(R)-x)\Theta_{1n}^{}(R,x)$ with $H(x^0(R)-x)$ the
Heaviside step function; 
$\widetilde{\Theta}_{1n}^{(2)}(R,x)$ and  $\Theta_{1x}^{(2)}(R,x)$  are the minimum and maximum
angles  (both in the range from 0 to $\pi$) between hb-arm 1 and $r$-axis with two 
hydrogen bonds formed simultaneously (for given $R,x$); 
$\Theta_{2n}^{(2)}(R,x,\Theta_1)$ and  $\Theta_{2x}^{(2)}(R,x,\Theta_1)$  are the
minimum and maximum angles 
(both in the range from 0 to $\pi$) that hb-arm 2 can have with the $r$-axis when two 
hydrogen bonds are formed simultaneously (for given $R,x$, and  
$\widetilde{\Theta}_{1n}^{}(R,x)\le\Theta_1\le\Theta_{1x}^{M}(R,x)$).
According to eqs.(S17) and (S3), the coefficient $k_{3(2,1)}(R,x)$ is thus given by
\beq k_{3(2,1)}(R,x)=\int_{\Theta_{1n}^{(3)}(R,x)}^{\Theta^{(3)}_{1x}(R,x)}
\frac{\mathrm{d}\Theta_1}{\Theta_{1x}^{M}(R,x)-\widetilde{\Theta}_{1n}^{}(R,x)} 
\int_{\Theta_{2n}^{(3)}(R,x,\Theta_1)}^{\Theta{2x}^{(3)}(R,x,\Theta_1)}
\frac{\mathrm{d}\Theta_2}{\Theta_{2x}^{}(R,x,\Theta_1)-\Theta_{2n}(R,x,\Theta_1)}. \eeq

For $R\ge R^0$ {\bf and} $x\in\;[0;x^0(R)]$, the angle $\Theta_{1n}^{(3)}(R,x)$ is obtained as the
solution of the equation 
\beq 
(1+\cos(\alpha))x_{+}^2+(-1+\cos(\alpha))y_{\pm}^2+(-1+\cos(\alpha))(z_{\pm}-(R+\eta+x))^2=0
\eeq 
with respect to $\Theta_1$ (recall that $x_{+}$, $y_{\pm}$, and $z_{\pm}$ are all functions of
$R,x$, and $\Theta_1$). If $0<R<R^0$ {\em or} $x\in\;\;]x^0(R),2\eta]$, the angle 
$$\Theta_{1n}^{(3)}(R,x)=0.$$ 
On the other hand, 
$$\Theta_{1x}^{(3)}(R,x)=\Theta_{1x}^{M}(R,x)$$ 
for any $R$ and $x$. 

Next, for $R\ge R^0$ {\bf and} $x\in\;[\eta;x^0(R)]$, 
\beq \Theta_{2x}^{(3)}(R,x,\Theta_1)=\left\{\begin{array}{ll} \Theta_{1x}^{M}(R,x) & \;\;\; 
\mbox{if}\;\;\Theta_1\in\;\;[\Theta_{1n}^{(3)}(R,x);\Theta_{1x}^{mM}(R,x)],\\
\pi-|\pi-(\alpha+\Theta_1)| & \;\;\; \mbox{otherwise},\end{array} \right. \eeq  
If $0<R<R^0$ {\em or} $x\in\;\;]x^0(R),2\eta]$, 
\beq \Theta_{2x}^{(3)}(R,x,\Theta_1)=\left\{\begin{array}{ll} \Theta_{1x}^{M}(R,x) & \;\;\; 
\mbox{if}\;\;\Theta_1\in\;\;[\Theta_{10}^{}(R,x);\Theta_{1x}^{mM}(R,x)],\\
\pi-|\pi-(\alpha+\Theta_1)| & \;\;\; \mbox{otherwise},\end{array} \right. \eeq  
 
Further, for $R\ge R^0$ {\bf and} $x\in\;[\eta;x^0(R)]$, 
\beq \Theta_{2n}^{(3)}(R,x,\Theta_1)=\left\{\begin{array}{ll} 
\frac{\pi}{2}-\arcsin\left(\frac{(Z_{2n}(R,x,\Theta_1)-Z_s)}{\eta}\right) & \;\;\; 
\mbox{if}\;\;\Theta_1\in\;\;[\Theta_{1n}^{(3)}(R,x);\Theta_{1x}^{mM}(R,x)],\\
|\alpha-\Theta_1| & \;\;\; \mbox{otherwise},\end{array} \right. \eeq  
where $Z_{3}^{(2)}=Z_{2n}(R,x,\Theta_1)$ is the $z$-coordinate of the tip of 
hb-arm 3 when the tip of hb-arm 2 is 
located on the sphere $S_{SHL}^{ib}$ and $Z_s$ are the $z$-coordinate of the selected molecules
$S$. 

For $R\ge R^0$ {\bf and} $x\in\;[x^0(R);x_{\mbox{\tiny sep}}(R)]$, 
\beq \Theta_{2n}^{(3)}(R,x,\Theta_1)=\left\{\begin{array}{ll} 
\frac{\pi}{2}-\arcsin\left(\frac{(Z_{2n}(R,x,\Theta_1)-Z_s)}{\eta}\right) & \;\;\; 
\mbox{if}\;\;\Theta_1\in\;\;[\Theta_{1low}^{(3)}(R,x);\Theta_{1up}^{(3)}(R,x)],\\
|\alpha-\Theta_1| & \;\;\; \mbox{otherwise},\end{array} \right. \eeq  
where $x_{\mbox{\tiny sep}}\equiv x_{\mbox{\tiny sep}}(R)$ is the $R$-dependent solution 
of a couple of simultaneous equations
$x_{+}(R,x,\Theta_1)=\eta\sin(\alpha/2)$ and $y_{\pm}(R,x,\Theta_1)=0$ 
with respect to $x$, and  $\Theta_{1low}^{(3)}(R,x)$ and $\Theta_{1up}^{(3)}(R,x)$ are the smaller
and larger solutions of the equation $x_{+}(R,x,\Theta_1)=\eta\sin(\alpha/2)$ 
(these solutions exist only at $x\le x_{\mbox{\tiny sep}}$ and at $x=x_{\mbox{\tiny sep}}$ they
degenerate into a single solution $\Theta_{1sep}^{(3)}(R)\equiv 
\Theta_{1low}^{(3)}(R,x_{\mbox{\tiny sep}}(R))=\Theta_{1up}^{(3)}(R,x_{\mbox{\tiny sep}}(R))$). 

For any $R\ge 0$ {\bf and} $x\in\;[x_{\mbox{\tiny sep}}(R);2\eta]$, 
\beq \Theta_{2n}^{(3)}(R,x,\Theta_1)= |\alpha-\Theta_1|\;\;\;\;\; 
\mbox{if}\;\;\Theta_1\in\;\;[0;\Theta_{1x}^{M}(R,x)], \eeq  

For $0<R<R^0$ {\em and} $x\in\;[\eta;x_{\mbox{\tiny sep}}(R)]$, 
\beq \Theta_{2n}^{(3)}(R,x,\Theta_1)=\left\{\begin{array}{ll} 
\frac{\pi}{2}-\arcsin\left(\frac{(Z_{2n}(R,x,\Theta_1)-Z_s)}{\eta}\right) & \;\;\; 
\mbox{if}\;\;\Theta_1\in\;\;[\Theta_{1low}^{(3)}(R,x);\Theta_{1up}^{(3)}(R,x)],\\
|\alpha-\Theta_1| & \;\;\; \mbox{otherwise},\end{array} \right. \eeq  

\subsubsection*{Coefficient $k_4$} 

Again, let us consider molecule $S$ in the SHL of a spherical 
hydrophobe of radius $R$ at a distance $x$ from its surface. 
The coefficient $k_4$ is  calculated by assuming that hb-arms 1,2, and 3 have already formed
bonds and taking into account that, if the water molecule is far enough from the hydrophobic surface 
(but still in the LHS), hb-arm 4 can still form a bond if its tip is 
{\em not} within the inner boundary of the hydrophobe SHL; besides, 
the angle  between any two of four hb-arms must be equal to $\alpha$. 
While the orientations of arms 1 and 2 are arbitrary  (with the angle between them equal to
$\alpha$), hb-arm 3 can have any of just two possible orientations determined by those of arms 
1 and 2, whereas the orientation of hb-arm 4 is uniquely determined by the orientations of hb-arms
1, 2, and 3. (Again, the orientations of arms 1 and 2 are eventually averaged assuming their  uniform
distributions. 

What is the probability $s'_{4(3,2,1)}\equiv s'_{4(3,2,1)}(x',\Theta_1,\Theta_2)$ that hb-arm 4
will form a bond as well? Let us denote  the angle  between arm $i$ ($i=1,2,3,4$) and axis $x'$ by
$\Theta_i$ and  introduce the same Cartesian coordinate system as above 
(in calculating $k_2(R,x)$) but with the origin coinciding with $S$. 
The  Cartesian coordinates of the tip of arm
$i\;\;(i=1,2,3,4)$ will be denoted by $x_i,y_i,z_i$.  First of all, in order for hb-arm 4 to form
a bond, molecule $S$ must be sufficiently far away from the inner boundary of the hydrophobe SHL
(represented the sphere $S_{SHL}^{ib}$). 
The minimum
distance $x_{min}$, at which this is possible, depends on $R$ and is equal to $x^0(R)$ defined
above and determined by eq.(S9). 

Further,  hb-arm 4 can form a bond with the same probability as in the bulk if the location of
molecule $S$ and orientation of its arms 1 and 2 (i.e., $x',\Theta_1,$ and $\Theta_2$)  are such
that the tip of arm 4 is not located within the sphere $S_{SHL}^{ib}$. 
Otherwise, hb-arm 4 cannot form a bond at all. Keeping in mind that
$s'_{4(3,2,1)}\equiv s'_{4(3,2,1)}(R,x,\Theta_1,\Theta_2)$, one can thus write 
\beq s'_{4(3,2,1)}=\left\{\begin{array}{ll}  b_{3(2,1)} & \;\;\; 
\mbox{if}\;\;\Theta_1\in\;\;\De_{1(4)}^{\Theta}\;\;\mbox{and}\;\;
\Theta_2\in\;\;\De_{2(4)}^{\Theta} ,\\
0 & \;\;\; \mbox{otherwise}.\end{array} \right. \eeq  
Here, $\De_{1(4)}^{\Theta}\equiv\De_{1(4)}^{\Theta}(R,x)$  is the range of angles 
(from 0 to $\pi$) between hb-arm 1 and $r$-axis with 
four hydrogen bonds formed  (for given $R$ and $x$) and  $\De_{2(4)}^{\Theta}\equiv\De_{2(4)}^{\Theta}(R,x,\Theta_1)$ 
is the range of angles  
(from 0 to $\pi$) that hb-arm 2 can have with the $r$-axis when four hydrogen
bonds form (assuming that  $\Theta_1\in\;\;\De_{1(4)}^{\Theta}(R,x)$ for given $R,x$).

The mean  probability $s_{4(3,2,1)}$ that a boundary molecule (i.e., molecule $S$)  forms a fourth
hydrogen bond once its hb-arms 1,2, and 3 have  already formed bonds can then be obtained by 
averaging $s'_{4(3,2,1)}$  over all  possible orientations  of its hb-arms 1 and 2, $\Theta_1$ and
$\Theta_2$  (assumed to be distributed uniformly):  
\beq s_{4(3,2,1)}=\int\limits_{\De_{1(4)}^{\Theta}}^{}
\frac{\mathrm{d}\Theta_1}{\Theta_{1x}^{M}(R,x)-\Theta_{1n}^{(3)}(R,x)} 
\int\limits_{\De_{1(4)}^{\Theta}}^{}
\frac{\mathrm{d}\Theta_2}{\Theta_{2x}^{(3)}(R,x,\Theta_1)-\Theta_{2n}^{(3)}(R,x,\Theta_1)}
s'_{4(3,2,1)}(R,x,\Theta_1,\Theta_2).\eeq
Thus, according to eqs.(S27) and (S3), the coefficient $k_4(R,x)$ is equal to zero 
for $\eta\le x\le x^0(R)$, 
whereas for $x^0(R)<x\le 2\eta$ it is given by
\beq k_{4(3,2,1)}(R,x)=\int\limits_{\De_{1(4)}^{\Theta}}^{}
\frac{\mathrm{d}\Theta_1}{\Theta_{1x}^{M}(R,x)-\Theta_{1n}^{(3)}(R,x)} 
\int\limits_{\De_{1(4)}^{\Theta}}^{}
\frac{\mathrm{d} \Theta_2}{\Theta_{2x}^{(3)}(R,x,\Theta_1)-\Theta_{2n}^{(3)}(R,x,\Theta_1)}. \eeq

For any $R\ge 0$ one can show that   
\beq \De_{1(4)}^{\Theta}=\left\{\begin{array}{ll} \; [0;\Theta_{1low}^{(3)}(R,x)]\;\;\bigcup \;\;
[\Theta_{1up}^{(3)}(R,x);\Theta_{1x}^{M}(R,x)]  &  
\mbox{if}\;\;x\in\;[x^0(R);x_{\mbox{\tiny sep}}(R)],\\ 
\; [0;\Theta_{1x}^{M}(R,x)] & \mbox{if}\;\;x\in\;[x_{\mbox{\tiny sep}}(R);2\eta].\end{array} 
\right. \eeq  

For any $R\ge 0$ {\bf and} $x\in\;[x_{\mbox{\tiny sep}}(R);2\eta]$, 
\beq \De_{2(4)}^{\Theta}=\left\{\begin{array}{ll} \; 
[\Theta_{2n}^{(4)}(\Theta_1);\Theta_{2x}^{(4)}(\Theta_1)]\hspace{1.5cm}   
\mbox{if}\;\;\;\;\Theta_1\in\;[0;\Theta_{10}(R,x)]\;\;\bigcup 
\;\; [\Theta_{1x}^{m}(R,x);\Theta_{1x}^{M}(R,x)],\\ 
\; [\Theta_{2n}^{(4)}(\Theta_1);\Theta_{1nx}^{(4)}(R,x)]\;\bigcup \;
[\Theta_{1xn}^{(4)}(R,x);\Theta_{1x}^{M}(R,x)] \hspace{0.5cm} 
\mbox{if}\;\;\;\;\Theta_1\in\;[\Theta_{10}(R,x);\Theta_{1x}^{m}(R,x)],\end{array} 
\right. \eeq  
where $\Theta_{2n}^{(4)}(\Theta_1)=|\alpha-\Theta_1|$ and 
$\Theta_{2x}^{(4)}(\Theta_1)=\pi-|\pi-(\alpha+\Theta_1)|$

For any $R\ge 0$ {\bf and} $x\in\;[x^0(R);x_{\mbox{\tiny sep}}(R)]$, 
\beq \De_{2(4)}^{\Theta}=\left\{\begin{array}{ll} \; 
[\Theta_{2n}^{(4)}(\Theta_1);\Theta_{2x}^{(4)}(\Theta_1)]\hspace{1.5cm}   
\mbox{if}\;\;\;\;\Theta_1\in\;[0;\Theta_{10}(R,x)]\;\;\bigcup 
\;\; [\Theta_{1x}^{m}(R,x);\Theta_{1x}^{M}(R,x)],\\  
\O \hspace{5.5cm} 
\mbox{if}\;\;\;\;\Theta_1\in\;[\Theta_{1low}^{(3)}(R,x);\Theta_{1up}^{(3)}(R,x);],\\ 
\; [\Theta_{2n}^{(4)}(\Theta_1);\Theta_{2nx}^{(4)}(R,x,\Theta_1)]\;\bigcup \;
[\Theta_{2xn}^{(4)}(R,x,\Theta_1);\Theta_{1x}^{M}(R,x)] \hspace{0.5cm} 
\mbox{otherwise}\;\;\;\,\end{array} 
\right. \eeq  
(the last condition ``otherwise" stands for $\Theta_1\in\;[\Theta_{10}(R,x);\Theta_{1low}^{(3)}(R,x)]\;\;\bigcup 
\;\; [\Theta_{1up}^{(3)}(R,x);\Theta_{1x}^{m}(R,x)]$. 

The angles 
$\Theta_{2nx}^{(4)}(R,x,\Theta_1)]$ and $\Theta_{2xn}^{(4)}(R,x,\Theta_1)]$ in eqs.(S31) and (S32)
are determined as 
\beq \Theta_{2nx}^{(4)}(R,x,\Theta_1)]=\left\{\begin{array}{ll} \;  
\arccos[z_{2}^{-}(R,x,\Theta_1)] &    
\mbox{if}\;\;\;\;|\alpha-\Theta_{1sep}^{(3)}(R)|=
\arccos[z_{2}^{-}(R,x_{\mbox{\tiny sep}}(R),\Theta_{1sep}^{(3)}(R))],\\ 
\; \arccos[z_{2}^{+}(R,x,\Theta_1)] & \mbox{otherwise} ,\end{array} 
\right. \eeq  
and 
\beq \Theta_{2xn}^{(4)}(R,x,\Theta_1)]=\left\{\begin{array}{ll} \;  
\arccos[z_{2}^{+}(R,x,\Theta_1)] &    
\mbox{if}\;\;\;\;\Theta_{1xn}^{(4)}(R,x,\Theta_1)]=\arccos[z_{2}^{-}(R,x,\Theta_1)],\\ 
\; \arccos[z_{2}^{-}(R,x,\Theta_1)] & \mbox{otherwise}.\end{array} 
\right. \eeq  
Here $z_{2}^{+}=z_{2}^{+}(R,x,\Theta_1)$ and $z_{2}^{-}=z_{2}^{-}(R,x,\Theta_1)$ 
are two solutions of the quadratic equation  
\beq
x_3\sqrt{1-(\cos(\alpha)/y_1-(z_1/y_1)z_2)^2-z_2^2}=\cos(\alpha)-
(\cos(\alpha)/y_1-(z_1/y_1)z_2)y_3-z_2z_3,
\eeq
with $y_1=-\sin\Theta_1,\, z_1=\cos\Theta_1,\, z_3=\cos\Theta_{1x}^{M}(x,R),\,
y_3=-(\cos(\alpha)/\sin\Theta_1)+z_3\cot\Theta_1,\, x_3=\sqrt{1-y_3^2-z_3^2}.$

\section*{References}
\begin{list}{}{\labelwidth 0cm \itemindent-\leftmargin} 
\item $(1)$ Sharp, K.A. {\it Curr. Opin. Struct. Biol.} {\bf 1991}, {\it 1}, 171-174. 
\item $(2)$ Blokzijl, W.; Engberts, J.B.F.N. {\it Angew. Chem. Int. Ed. Engl.} {\bf 1993}, {\it 32}, 1545-1579.
\item $(3)$ Soda, K. {\it Adv. Biophys.} {\bf 1993}, {\it 29}, 1-54. 
\item $(4)$ Paulaitis, M.E.; Garde, S.; Ashbaugh, H.S.    
{\it Curr. Opin. Colloid Interface Sci.} {\bf 1996}, {\it 1}, 376-383.
\item $(5)$ Ghelis, C.; Yan, J. {\it Protein Folding}; Academic Press: New York, 1982.
\item $(6)$ Kauzmann, W.  {\it Adv.Prot.Chem.} {\bf 1959}, {\it 14}, 1-63.
\item $(7)$ Ben-Naim, A. {\it J.Biomol.Struct.Dyn.} {\bf 2012}, {\it 30}, 113-124.
\item $(8)$ Ashbaugh, H.S.; Truskett, T.M.; Debenedetti, P. 
{\it Phys.Chem.Chem.Phys.} {\bf 2002}, {\it 116}, 2907-2921.
\item $(9)$ Widom, B.; Bhimulaparam, P.; Koga, K.  {\it Phys.Chem.Chem.Phys.} {\bf 2003}, {\it 5}, 3085-3093.
\item $(10)$ Ball, P. {\it Chem.Rev.} {\bf 2008}, {\it 108}, 74-108.
\item $(11)$ Berne B.J.; Weeks J.D.; Zhou R.  {\it Annu.Rev.Phys.Chem.} {\bf 2009}, {\it  60}, 85-103.
\item $(12)$ Bernal, J.D.; Fowler, R.H. {\it J.Chem.Phys.} {\bf 1933},  {\it 1},  515-548.
\item $(13)$ Koizumi, M.; Hirai, H.; Onai, T.; Inoue, K.; Hirai, M. {\it J.Appl.Cryst.} {\bf 2007}, {\it 40}, 
175-178. 
\item $(14)$ Southall, N.T.; Dill, K.A.  {\it J.Phys.Chem. B}, {\bf 2000}, {\it 104}, 1326-1331.
\item $(15)$ Rajamani, S.; Truskett, T.M.; Garde, S.{\it Proc.Natl.Acad.Sci.USA}  {\bf 2005}, {\it 102}, 
9475-9480.
\item $(16)$ Chandler, D. {\it Nature} {\bf 2005}, {\it 437}, 640-7.
\item $(17)$ Watanabe, K.; Andersen, H.C.  {\it J.Phys.Chem.} {\bf 1986}, {\it 90}, 795-802.
\item $(18)$ Pangali, C.; Rao,M.; Berne, B.J. {\it J.Chem.Phys.} {\bf 1979}, {\it 71}, 2982-90.
\item $(19)$ Pratt, L.R.; Chandler, D.  {\it J.Chem.Phys.} {\bf 1977}, {\it 67}, 3683-3704.
\item $(20)$ Lee,C.Y.; McCammon, J.A.; Rossky, P.J. {\it J. Chem. Phys.} {\bf 1984}, {\it 80}, 4448-55.
\item $(21)$ Stillinger, F.H. {\it J.Solut.Chem.} {\bf 1973}, {\it 2}, 141-58.
\item $(22)$ Pratt, L.R. {\it Annu. Rev. Phys. Chem.} {\bf 2002}, {\it 53}, 409-36
\item $(23)$ Kuipers, J.; Blokhius, E.M.  {\it J. Chem. Phys.} {\bf 2009}, {\it 131}, 044701..
\item $(24)$ Meng, E.C.; Kollman, P.A. {\it J.Phys.Chem.} {\bf 1996}, {\it 110}, 11460-11470.
\item $(25)$ Silverstein, K.A.T.; Haymet, A.D.J.; Dill, K.A.  
{\it J.Chem.Phys.} {\bf 1999}, {\it 111}, 8000-8009.
\item $(26)$ Chaplin, M.F. in {\em Water of Life: The unique properties of H$_2$0}, edited by  
R.M.Lynden-Bell, S.C.Morris, J.D.Barrow, J.L.Finney, and C.Harper, (CRC Press, Boca Raton,
2010), p.69. 
\item $(27)$ Djikaev, Y.S.; Ruckenstein, E.  
{\it Curr Opin Colloid Interface Sci.} {\bf 2011}, {\it 16}, 272; doi:10.1016/j.cocis.2010.10.002 
\item $(28)$ Luzar, A.; Svetina, S.; Zeks, B. {\it J.Chem.Phys.} {\bf 1985}, {\it 82}, 5146.
\item $(29$ Zangi, R.; Berne, B.J. {\it J.Phys.Chem.B} {\bf 2008}, {\it 112}, 8634-8644.
\item $(30)$ Evans, R.  in {\it Fundamentals 
of inhomogeneous fluids}, ed. D. Henderson; Marcel Dekker: New York, 1992. 
\item $(31)$ Sullivan, D.E. {\it Phys.Rev. B} {\bf 1979}, {\it 20}, 3991-4000. 
\item $(32)$ Tarazona, P.; Evans, R. {\it Mol. Phys.} {\bf 1983}, {\it 48}, 799-831.
\item $(33)$ Nakanishi, H.; Fisher, M.E., {\it Phys.Rev.Lett.} {\bf 1982}, {\it 49}, 1565-1568.
\item $(34)$ Ruckenstein, E.; Djikaev, Y.S. {\it J. Phys. Chem. Lett.} {\bf 2011}, {\it 2}, 1382-1386. 
\item $(35)$ Djikaev, Y.S.; Ruckenstein, E. {\it J. Phys. Chem. B} {\bf 2012}, {\it 116 }, 2820-2830.
\item $(36)$ Tarazona,P.  {\it Phys.Rev.A} {\bf 1985}, {\it 31}, 2672-2679; {\it 32}, 3140 (erratum).
\item $(37)$ Tarazona,P.; Marconi,U.M.B.; Evans, R. {\it Mol.Phys.} {\bf 1987},{\it 60}, 573-579.
\item $(38)$ Carnahan, N.F.; Starling, K.E. {\it J. Chem. Phys.}, {\bf 1969}, {\it 51}, 635-6.
\item $(39)$ Weeks, J.D.; Chandler, D.; Anderson, H.C. {\it J. Chem. Phys.}, {\bf 1971}, {\it 54}, 5237-47.
\item $(40)$ Patel, L.J.; Varilly, P.; Chandler, D. {\it J.Phys.Chem.B} {\bf 2010}, {\it 114}, 1632-1637.
\item $(41)$ Oleinikova, A; Brovchenko, I. {\it J.Phys.Chem.B} {\bf 2012}, {\it 116}, 14650-14659.

\end{list}

\newpage 
\subsection*{Captions} 
to  Figures 1 through 5 of the manuscript 
{\sc
``Probabilistic approach to the length-scale dependence of the effect of water hydrogen bonding on 
hydrophobic hydration"
}  by {\bf Y. S. Djikaev} and  {\bf E. Ruckenstein}.  
\subsubsection*{}
Figure 1. A water molecule in the surface hydration layer (SHL) of a spherical hydrophobic particle of radius $R$. 
The inner boundary of
the SHL is a sphere 
of radius 
$R+\eta$,  while the
outer (closer to the bulk water) boundary is shown as the sphere of radius $R+2\eta $.  
The molecule, shown as disk $S$, is located at the distance $x$  from the hydrophobe surface. 
Two hb-arms of the molecule (arms 1 and 2) are in the plane of the  Figure, while arms
3 and 4 are located out of the Figure plane  (one of them under, the other above).  The tips
of hb-arms are shown as empty circles. The angle between any two hb-arms is $\alpha$. The
origin $O$ of the Cartesian coordinate system lies coincides with the center of the spherical hydrophobe,. 
The angle between hb-arm $i\;\;(i=1,2,3,4)$ and axis $x$ is denoted by $\Theta_i$.
\vspace{0.3cm}\\
Figure 2.  
The function $n_s(R,x)$ for a spherical hydrophobe in liquid water at
temperature $T=293.15$ K (with $n_b=3.65$): a) $n_s$ 
vs $\xi\equiv (x/\eta-1)$ for various $R$'s; b) 
$n_s$ vs $R$ at different $x$'s.
\vspace{0.3cm}\\ 
Figure 3. 
The water density profiles near a hydrophobe of 
radius $R$ (for $R/\eta=1,3,5,7,10,15,30,100$) 
with $\eps_{\mbox{\tiny wp}}/\eps_{\mbox{\tiny ww}}=0.75$  at $T=293.15$ K and 
$\mu=-11.5989$ $k_BT$. 
\vspace{0.3cm}\\ 
Figure 4. 
The water density profiles near a hydrophobe of 
radius $R$ at $T=293.15$ K and $\mu=-11.5989$ $k_BT$, with a) $R/\eta=1$ and b) $R/\eta=1$. 
Three curves each of Figs.4a and 4b correspond to different  
$\eps_{\mbox{\tiny wp}}/\eps_{\mbox{\tiny ww}}$, 
with $\eps_{\mbox{\tiny wp}}/\eps_{\mbox{\tiny ww}}=0.6,0.75$, and $1.2$ for curves from bottom to top. 
\vspace{0.3cm}\\
Figure 5.  
The grand canonical free energy of hydration of a spherical 
hydrophobe of radius $R$. The dimensionless 
$\overline{\De \Omega_{\mbox{\tiny }}}\equiv \De \Omega_{\mbox{\tiny }}/(k_BT(4\pi R^2/\eta^2))$
is plotted: a) vs $R/\eta$ for particles with 
$\eps_{\mbox{\tiny wp}}/\eps_{\mbox{\tiny ww}}=0.75$; 
b) vs  
$\eps_{\mbox{\tiny wp}}/\eps_{\mbox{\tiny ww}}$ for various $R$'s  
(the curves are for $R/\eta=1,\,3,\,7,\,15,$, and $30$ from bottom to top)

\newpage
\begin{figure}[htp]
\begin{center}\vspace{1cm}
\includegraphics[width=8.9cm]{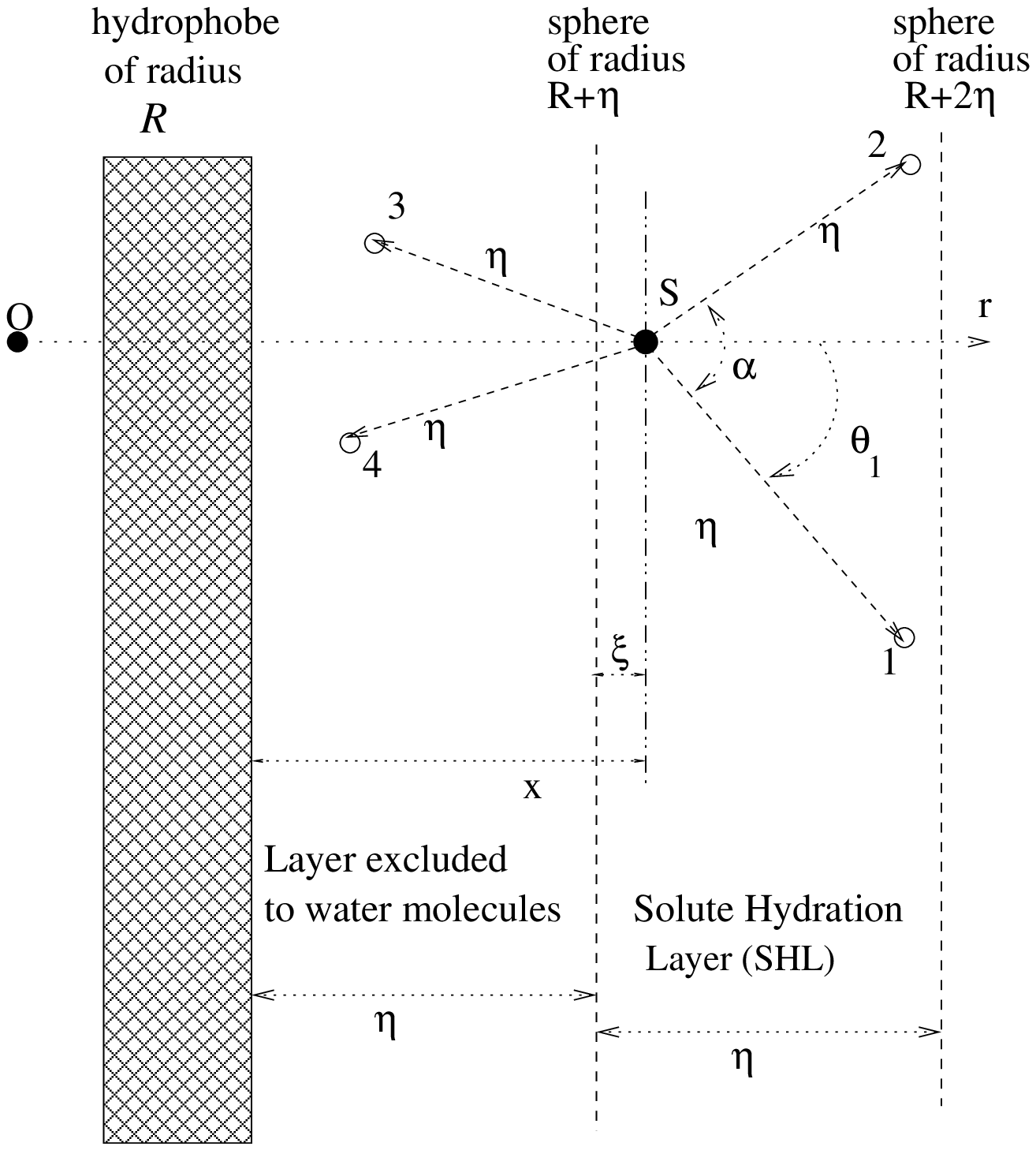}\\ [3.7cm]
\caption{\small }
\end{center}
\end{figure} 

\newpage
\begin{figure}[htp]
\begin{center}\vspace{1cm}
\includegraphics[width=8.9cm]{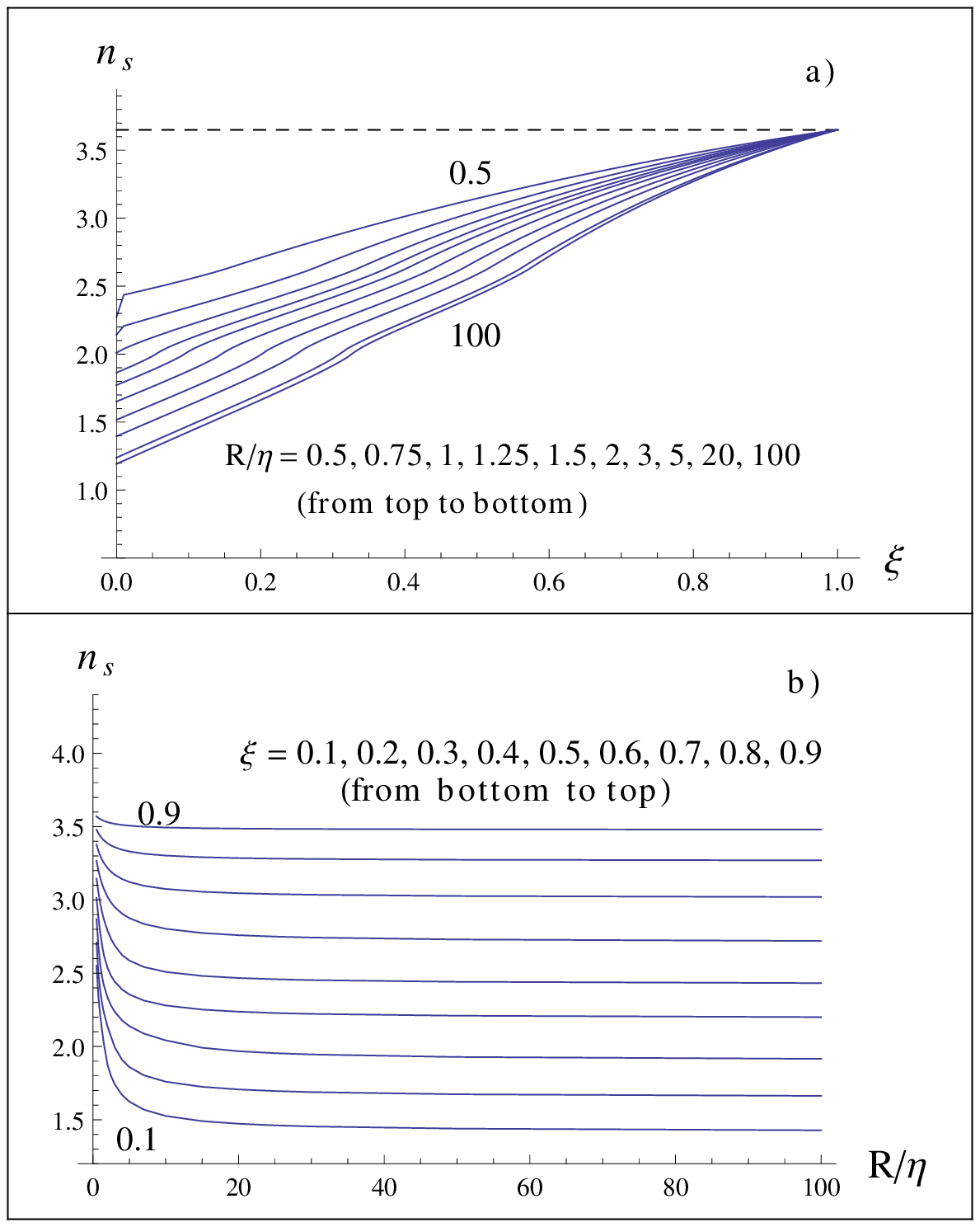}\\ [3.7cm]
\caption{\small }
\end{center}
\end{figure} 

\newpage
\begin{figure}[htp]
\begin{center}\vspace{1cm}
\includegraphics[width=8.9cm]{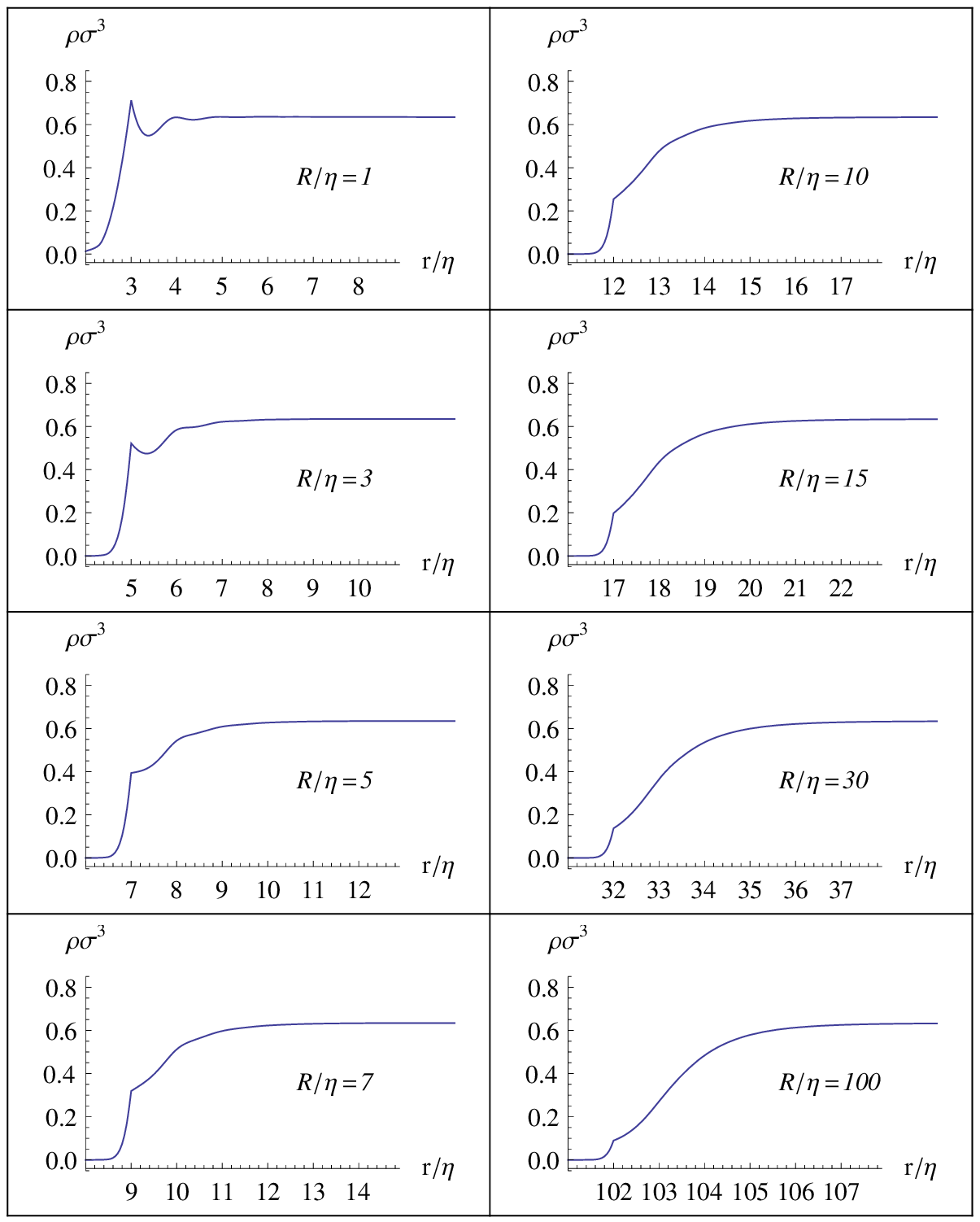}\\ [3.7cm]
\caption{\small }
\end{center}
\end{figure} 

\newpage
\begin{figure}[htp]
\begin{center}\vspace{-1cm}
\includegraphics[width=8.3cm]{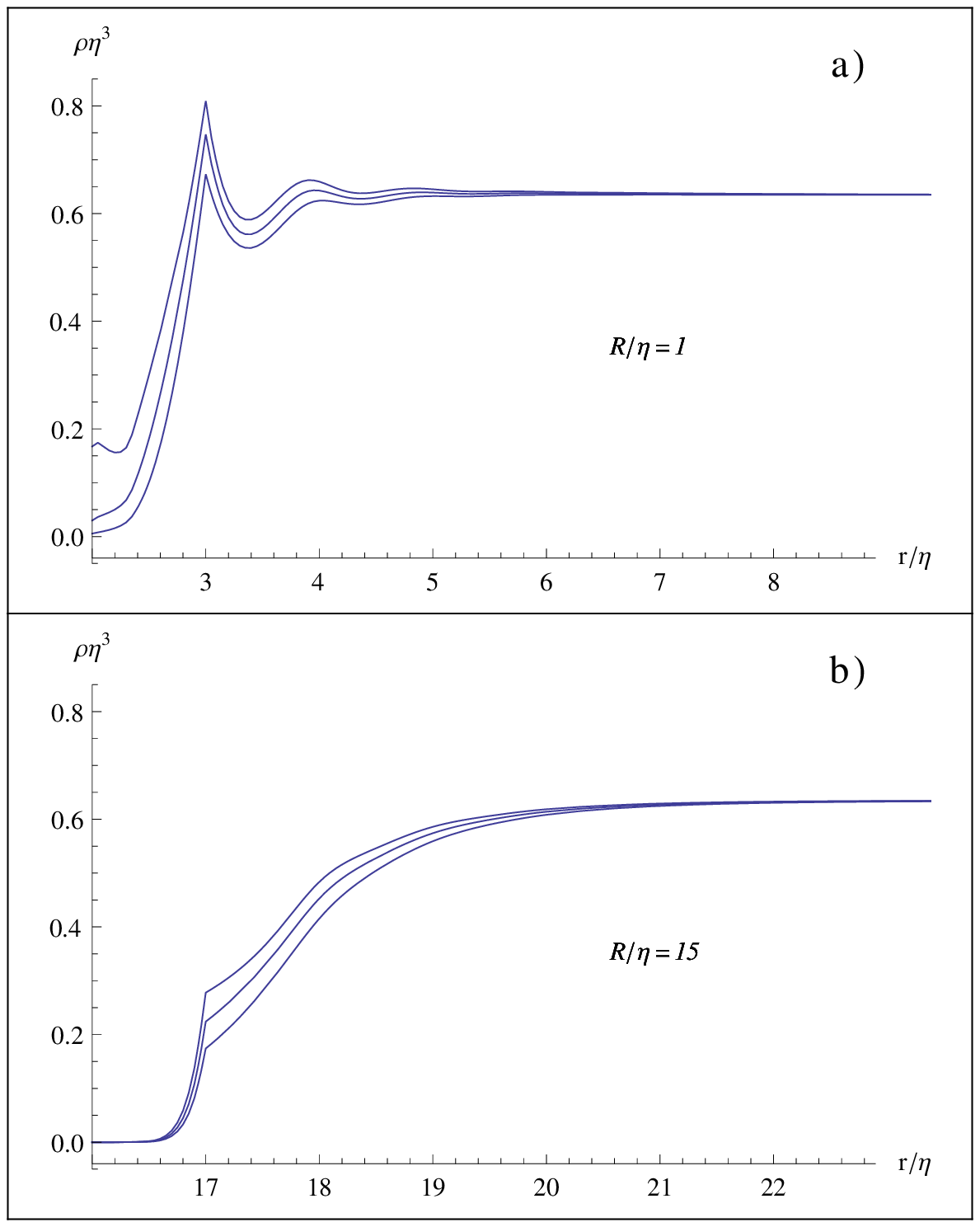}\\ [3.7cm]
\caption{\small }
\end{center}
\end{figure} 

\newpage
\begin{figure}[htp]
\begin{center}\vspace{1cm}
\includegraphics[width=8.9cm]{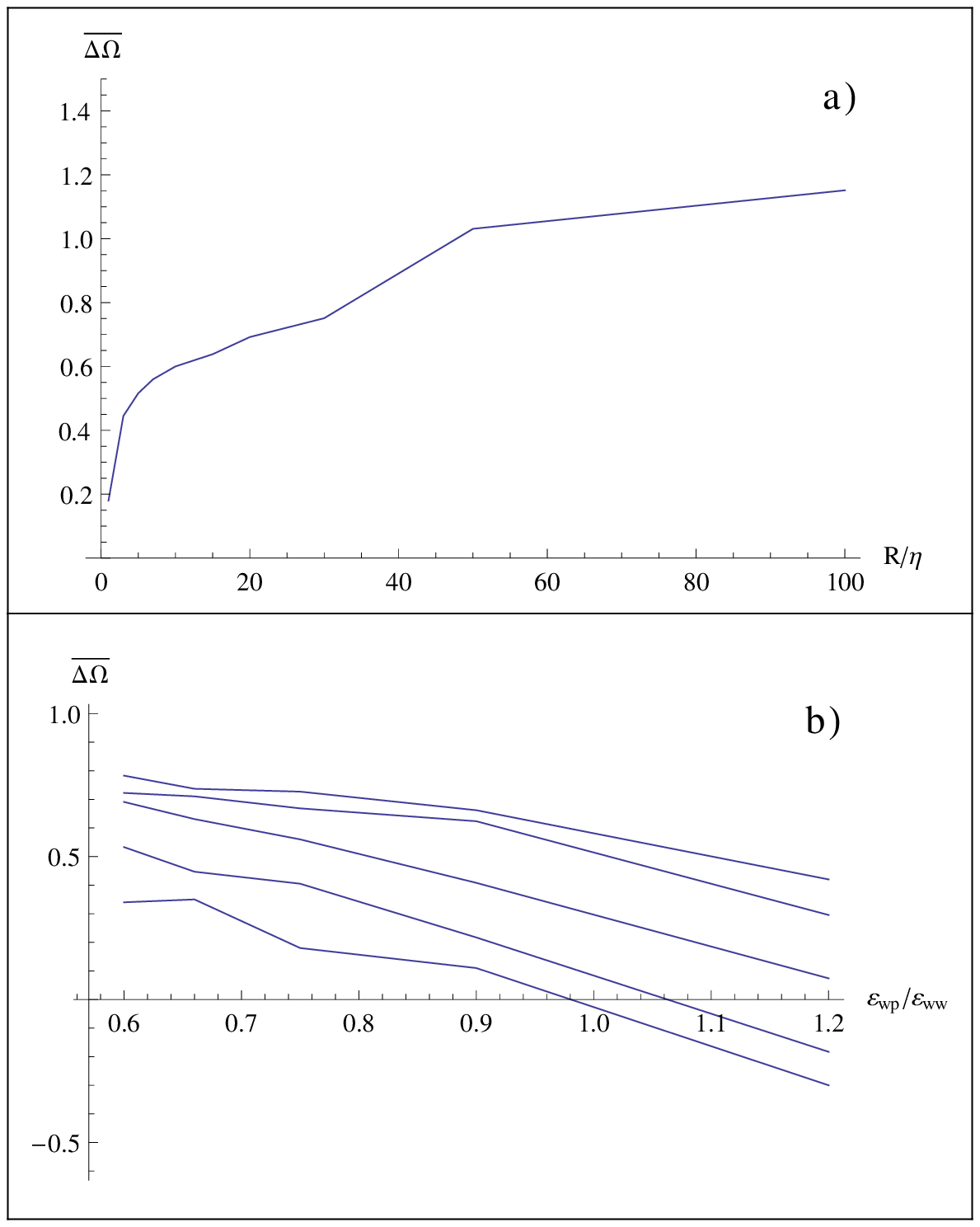}\\ [3.7cm]
\caption{\small }
\end{center}
\end{figure} 

\newpage
\begin{figure}[h]
\begin{center}\vspace{-0.2cm}
\includegraphics[width=8.3cm]{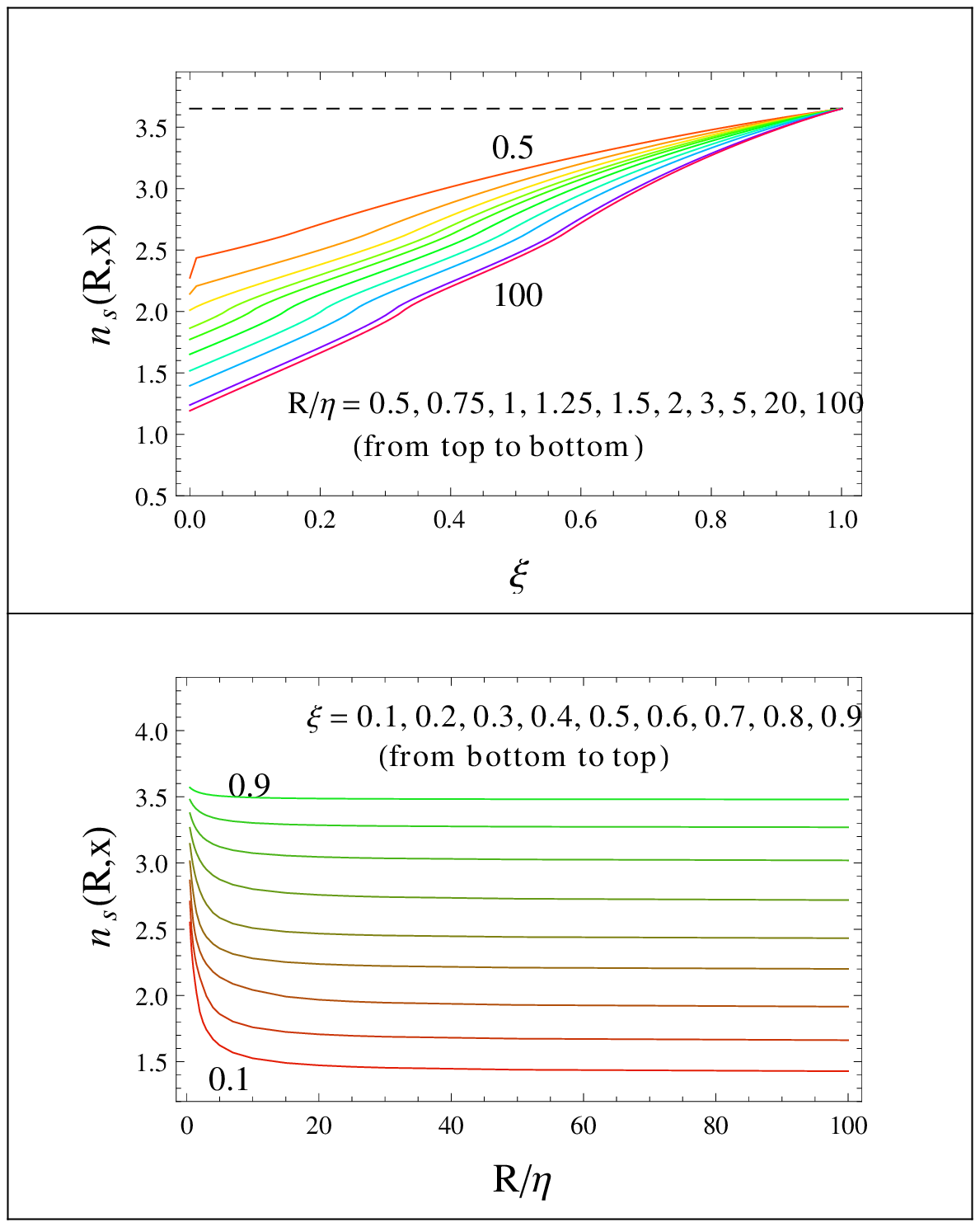}\\ [1.5cm]
\vspace{2.0cm}
\end{center}
\centerline{\Large TOC Graphic } 
\end{figure} 
\end{document}